\documentclass[a4paper,fleqn]{cas-dc}

\usepackage[authoryear,longnamesfirst]{natbib}

\def\tsc#1{\csdef{#1}{\textsc{\lowercase{#1}}\xspace}}
\tsc{WGM}
\tsc{QE}
\usepackage{amsmath}
\usepackage{amsfonts}				%% ドイツ文字など
\usepackage{mathrsfs} 				%% 花文字

\usepackage{bm}					%% 数式で太字の斜体（ボールドイタリック）を使う
\usepackage{textcomp}				%% 登録商標

\usepackage[hang,small,bf]{caption}
\usepackage[subrefformat=parens]{subcaption}
\usepackage{wrapfig}

\usepackage{color}

\usepackage{caption}
\usepackage{subcaption}

\usepackage{adjustbox}	%列にまたがるサイズの表がある場合

\begin{document}
\let\WriteBookmarks\relax
\def\floatpagepagefraction{1}
\def\textpagefraction{.001}

\shorttitle{%Reliability of Posterior Probability Estimation \\ under Prior Probability Bias \\ in Medical Image Training Data

Reliability Assessment Framework for Pathological Cell Image Classification}    

\shortauthors{T. Tachibana and T. Nagasaka}  

\title [mode = title]{%Reliability of Posterior Probability Estimation \\ under Prior Probability Bias \\ in Medical Image Training Data

Reliability Assessment Framework Based on Feature Separability for Pathological Cell Image Classification under Prior Bias}  



%authors


\author[1]{Takaaki Tachibana}%[<options>]

%\cormark[1]


\ead{tachi93@med.kobe-u.ac.jp}


\credit{Investigation, Formal analysis, Writing – original draft}

\affiliation[1]{organization={Division of Gastrointestinal Surgery, Department of Surgery,
Kobe University Graduate School of Medicine},
            addressline={7-5-2 Kusunoki-cho, Chuo-ku,}, 
            city={Kobe},
            citysep={}, % Uncomment if no comma needed between city and postcode
            postcode={650-0017}, 
            state={Hyogo},
            country={Japan}}

\author[2,1]{Toru Nagasaka}%[]

\cormark[1]

% Footnote of the second author

% Email id of the second author
\ead{toru-ngy@umin.ac.jp}


\credit{Conceptualization, Methodology, Software, Data curation, Visualization, Validation, Writing – review and editing}

\affiliation[2]{organization={Association of Medical Artificial Intelligence Curation},
            addressline={505, Sakae Members Office Building, 4-16-8 Sakae, Naka-ku}, 
            city={Nagoya},
            citysep={}, % Uncomment if no comma needed between city and postcode
            postcode={460-0008}, 
            state={Aichi},
            country={Japan}}

\author[1]{Yukari Adachi}%[]
\credit{Investigation, Data curation}

\author[1]{Hiroki Kagiyama}%[]
\credit{Investigation, Data curation}

\author[1]{Ryota Ito}%[]
\credit{Investigation, Data curation}

\author[3]{Mitsugu Fujita}%[]

% Footnote of the second author
%\fnmark[3]

\credit{Writing – review and editing}

\affiliation[3]{organization={Center for Medical Education and Clinical Training, Kindai University Faculty of Medicine},
            addressline={1-14-1 Miharadai, Minami-ku}, 
            city={Sakai},
            citysep={}, % Uncomment if no comma needed between city and postcode
            postcode={590-0111}, 
            state={Osaka},
            country={Japan}}

\author[4,1]{Kimihiro Yamashita}%[]

\affiliation[4]{organization={Department of Biophysics, Kobe University Graduate School of Health Sciences},
            addressline={7-10-2 Tomogaoka, Suma-ku}, 
            city={Kobe},
            citysep={}, % Uncomment if no comma needed between city and postcode
            postcode={654-0142}, 
            state={Hyogo},
            country={Japan}}

\credit{Resources, Project administration}

\author[1]{Yoshihiro Kakeji}%[]

\credit{Supervision}

\cortext[1]{Corresponding author}

\fntext[1]{}


\begin{abstract}
\relax%		%この何もしないコマンドを入れるとElsevierのスタイルのabstract環境において、Backgroundの太字ができるようになる
\textbf{Background and objective:} Prior probability shift between training and deployment datasets poses significant challenges for deep learning-based medical image classification. While standard correction methods reweight posterior probabilities to account for prior bias, their effectiveness varies unpredictably across applications. We developed a reliability assessment framework to determine when prior correction improves versus degrades classification performance in pathological cell image analysis.
\textbf{Methods:} We analyzed 303 colorectal cancer specimens with CD103/CD8 immunostaining, generating 185,432 annotated cell images across 16 cell types. ResNet models were trained under varying prior bias conditions (bias ratios 1.1--20×). We quantified feature separability using cosine similarity-based likelihood quality scores, measuring intra-class versus inter-class similarities in learned feature spaces. Multiple linear regression, ANOVA, and generalized additive models (GAMs) evaluated relationships between feature separability, prior bias magnitude, sample adequacy, and classification performance (F1 scores).
\textbf{Results:} Feature separability emerged as the dominant performance determinant ($\beta = 1.650$, $p < 0.001$), with 412-fold greater influence than prior bias ($\beta = 0.004$, $p = 0.018$). GAM analysis achieved strong predictive power ($R^2 = 0.876$), confirming predominantly linear relationships. A likelihood quality threshold of 0.294 effectively distinguished cases requiring correction, with ROC analysis demonstrating practical utility (AUC = 0.610). Cell types with quality scores exceeding 0.5 demonstrated reliable classification without correction, while those below 0.3 consistently required intervention for acceptable performance.
\textbf{Conclusion:} Feature extraction quality, not prior bias magnitude, determines when correction procedures benefit classification performance. Our framework provides quantitative guidelines for selective correction application, enabling resource-efficient deployment while maintaining diagnostic accuracy. This approach establishes principled decision-making for bias correction in clinical AI systems.
\end{abstract}


%\item 
%\item 
%\item 

%\nocite{*}

% Keywords

\maketitle

% Main text

\section{Introduction}

%We consider deep learning-based cell classification where the goal is to distinguish between multiple cell types in medical images. Let $Y \in \{1, 2, ..., K\}$ denote the cell type label, where $K$ is the total number of cell types. In many practical scenarios, the prior probability of observing cell type $c$ in the training dataset $\mathcal{D}$, denoted as $\hat{P}_{\mathcal{D}}(Y = c)$, differs from the true prior probability $P^*(Y = c)$ in the actual clinical population $\mathcal{P}$. Such prior probability shift is a frequently occurring problem in medical image analysis that can significantly impact model performance when deploying trained models to real-world clinical settings \citep{he2009learning, dockes2021preventing, guan2021domain}.

We consider deep learning-based cell classification where the goal is to distinguish between multiple cell types in medical images. In this setting, we aim to classify cells into one of $K$ distinct cell types, where each cell is assigned a label $Y \in \{1, 2, ..., K\}$ corresponding to its type. A critical challenge in medical image analysis arises when there is a mismatch between the distribution of cell types in the training dataset and the actual clinical population. Specifically, the prior probability of observing cell type $c$ in the training dataset $\mathcal{D}$, denoted as $\hat{P}_{\mathcal{D}}(Y = c)$, often differs from the true prior probability $P^*(Y = c)$ in the actual clinical population $\mathcal{P}$. Such prior probability shift is a frequently occurring problem in medical image analysis that can significantly impact model performance when deploying trained models to real-world clinical settings \citep{he2009learning, dockes2021preventing, guan2021domain}.

Several approaches have been proposed to address this prior probability shift problem, including iterative expectation-maximization methods \citep{saerens2002adjusting}, confusion matrix-based Black Box Shift Estimation \citep{lipton2018detecting}, and neural network-based Label Transformation Frameworks \citep{guo2020ltf}. Among these, the most commonly used standard approach applies prior probability correction through direct reweighting. Given a cell image $\mathbf{x}$, the corrected posterior probability for cell type $c$ is computed as:
\begin{equation}
P_{\text{corrected}}(Y = c | \mathbf{x}) = \frac{P_\theta(Y = c | \mathbf{x}) \cdot w_c}{\sum_{j=1}^{K} P_\theta(Y = j | \mathbf{x}) \cdot w_j}
\end{equation}
where $w_c = \frac{P^*(Y = c)}{\hat{P}_{\mathcal{D}}(Y = c)}$ represents the correction weight for cell type $c$, and $P_\theta(Y = c | \mathbf{x})$ is the posterior probability of cell type $c$ as estimated by the trained deep learning model with parameters $\theta$.

However, our preliminary analysis reveals that this correction is not universally beneficial for cell classification tasks. While some studies assume that prior correction should always improve performance when prior shift exists \citep{azizzadenesheli2019regularized}, empirical evidence suggests that the effectiveness of correction depends critically on model and data characteristics that have not been systematically investigated \citep{kull2019beyond, alexandari2020maximumlikelihoodbiascorrectedcalibration, bonab2026deep}. 
In particular, factors specific to cell classification—such as the quality of learned morphological features, the degree of class imbalance among cell types, and the intra-class heterogeneity of cellular appearance—may determine whether prior correction improves or degrades classification accuracy.

To systematically investigate when prior correction is effective, we first establish the theoretical foundation underlying posterior probability estimation in deep learning models.

According to Bayes' theorem, the posterior probability is formulated as:
\begin{equation}
P(C_i|\mathbf{x}) = \frac{P(\mathbf{x}|C_i) \cdot P(C_i)}{P(\mathbf{x})} = \frac{P(\mathbf{x}|C_i) \cdot P(C_i)}{\sum_{j=1}^{K} P(\mathbf{x}|C_j) \cdot P(C_j)}
\end{equation}

where:
\begin{itemize}
\item $P(C_i|\mathbf{x})$: posterior probability of class $C_i$ given image $\mathbf{x}$
\item $P(\mathbf{x}|C_i)$: likelihood of image $\mathbf{x}$ under class $C_i$
\item $P(C_i)$: prior probability of class $C_i$
\item $K$: total number of classes
\end{itemize}

Deep learning models $f_\theta$ can typically be decomposed into a feature extraction part $\Phi_\theta$ and a classification part $g_\theta$:
\begin{equation}
f_\theta(\mathbf{x}) = g_\theta(\Phi_\theta(\mathbf{x}))
\end{equation}

where $\Phi_\theta(\mathbf{x}) \in \mathbb{R}^d$ is a $d$-dimensional feature vector. By the chain rule of probability, the original likelihood can be decomposed as:
\begin{equation}
P^*(\mathbf{x}|C_i) = P^*(\Phi_\theta(\mathbf{x})|C_i) \cdot P^*(\mathbf{x}|\Phi_\theta(\mathbf{x}), C_i)
\end{equation}

This decomposition separates the likelihood into two components: the probability of the feature vector given the class, and the probability of the original image given both the feature vector and the class.

The effectiveness of prior correction is theoretically related to the concept of feature sufficiency. Ideally, when the feature vector captures all class-relevant information, the following conditional independence should approximately hold:
\begin{equation}
P^*(\mathbf{x}|\Phi_\theta(\mathbf{x}), C_i) \approx P^*(\mathbf{x}|\Phi_\theta(\mathbf{x}))
\end{equation}

This implies that given the feature vector $\Phi_\theta(\mathbf{x})$, the original image $\mathbf{x}$ and class label $C_i$ are approximately conditionally independent:
\begin{equation}
\mathbf{x} \perp C_i \mid \Phi_\theta(\mathbf{x})
\end{equation}

When this condition approximately holds, the posterior probability depends primarily on the feature vector:
\begin{align}
P^*(C_i|\mathbf{x}) &\approx P^*(C_i|\Phi_\theta(\mathbf{x}))
\end{align}

Under such conditions, high-quality likelihood estimation in the feature space may reduce the need for explicit prior correction.
From a theoretical perspective, the effectiveness of prior correction is closely related to the quality of likelihood estimation in the learned feature space. When deep learning models achieve high-quality feature representations that capture all class-relevant information (approaching the ideal condition of feature sufficiency), the need for prior correction may diminish or even become counterproductive. Conversely, when feature representations are suboptimal, prior correction may provide substantial benefits.

This raises several critical research questions: Under what conditions does prior correction improve classification performance? How can we predict when correction will be beneficial versus harmful? What model and data characteristics determine correction effectiveness?

To address these questions, we conduct a systematic empirical investigation examining the relationship between correction effectiveness and three key factors: (1) the quality of learned feature representations, assessed through feature space separability analysis; (2) the magnitude of prior probability shift between training and true conditions; and (3) the adequacy of training samples for reliable model estimation.

Our research contributes to the field in several ways. First, we provide the first systematic analysis of when prior correction is beneficial versus harmful in deep learning-based medical image classification. Second, we develop predictive models that can determine correction effectiveness based on readily observable model and data characteristics. Third, we establish practical guidelines for practitioners to decide when to apply prior correction in real-world deployment scenarios.

The results of this investigation will inform more reliable deployment strategies for deep learning models in clinical settings where training and deployment conditions may differ substantially. Our approach advances the field by developing a systematic framework for predicting correction effectiveness based on model and data characteristics, enabling principled decision-making for optimal intervention strategies in clinical deployment scenarios. This work establishes when and how to apply correction methods most effectively, ultimately contributing to more robust and reliable medical AI systems.

\section{Materials}

\subsection{Patients and Samples}

This study included tissue specimens from 303 colorectal cancer patients who underwent biopsy or surgery, with or without neoadjuvant chemoradiotherapy. Both biopsy and surgical samples were formalin-fixed and paraffin-embedded, and were subjected to dual immunostaining for CD103 and CD8. The staining process was performed according to established protocols to ensure consistency and reliability \citep{ohno2024tumor}.

\begin{figure}[h]
\begin{minipage}[t]{0.48\linewidth}
\includegraphics[width=1.0\columnwidth]{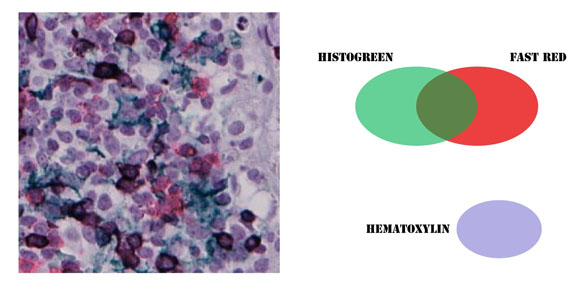}
\centering %\scriptsize{CD103/CD8 IHC}
%\label{fig:original}
\end{minipage}
\begin{minipage}[t]{0.48\linewidth}
\includegraphics[width=1.0\columnwidth]{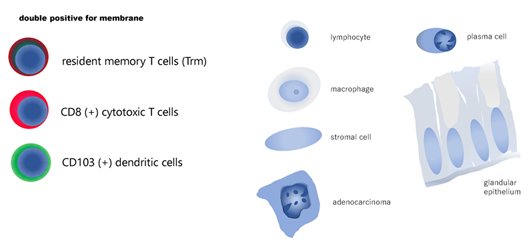}
\centering %\scriptsize{CD103/CD8 double stain}
%\label{fig:adenocarcinoma}
\end{minipage}

\caption{CD103/CD8 double stain}
\label{fig:CD103_CD8}
\end{figure}

\subsection{Data Annotation Methodology Using Cu-Cyto\textsuperscript{\tiny{\textregistered}} Viewer}
Cell-level annotation was performed using the Cu-Cyto\textsuperscript{\tiny{\textregistered}} Viewer, a specialized annotation tool designed for precise cellular identification in histopathological images \citep{abe2023deep}. The annotation workflow employed a human-AI collaborative approach to ensure comprehensive cellular coverage and annotation accuracy across all tissue sections.

\begin{figure}[htbp]
\centering
\begin{subfigure}[b]{0.48\columnwidth}
    \centering
    \includegraphics[width=\columnwidth]{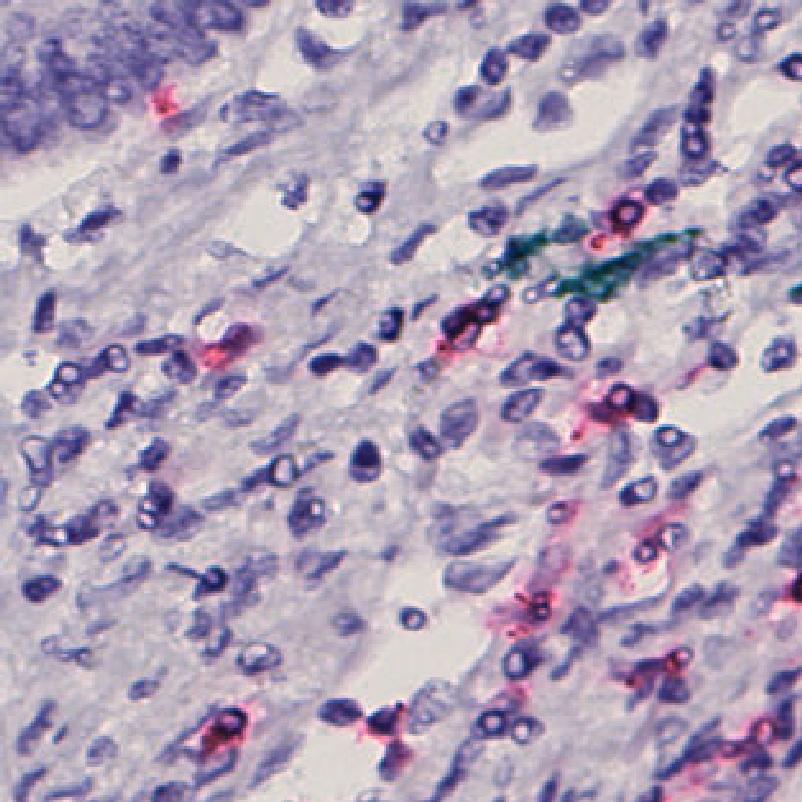}
    \caption{Original tissue section without annotations showing various cellular populations and tissue architecture.}
    \label{fig:before_annotation}
\end{subfigure}
\hfill
\begin{subfigure}[b]{0.48\columnwidth}
    \centering
    \includegraphics[width=\columnwidth]{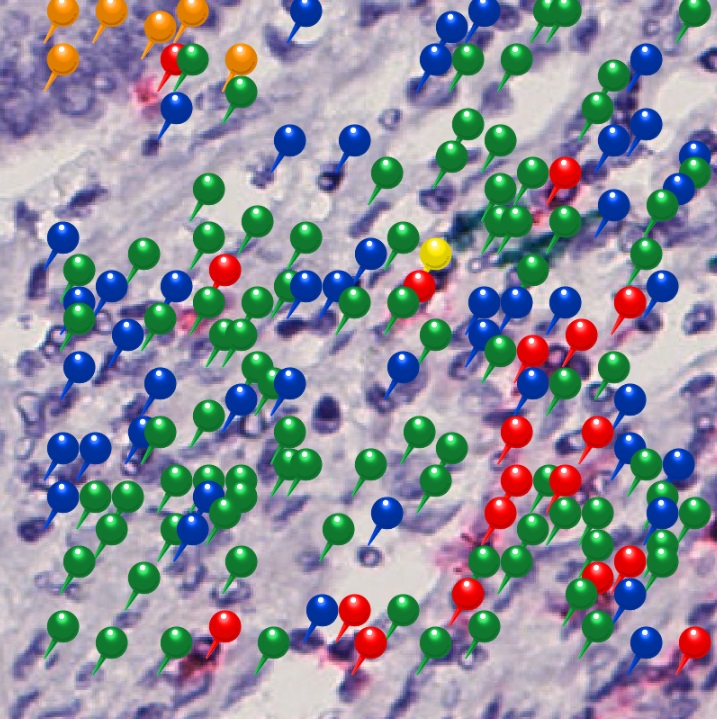}
    \caption{Comprehensive cellular annotations using colored markers to distinguish different cell types.}
    \label{fig:after_annotation}
\end{subfigure}
\caption{Cu-Cyto\textsuperscript{\tiny{\textregistered}} Viewer annotation workflow demonstration. (a) Unannotated histological section displaying cellular morphology and tissue organization. (b) Fully annotated section with color-coded markers identifying individual cells across different cellular populations. Each colored marker represents a distinct cell type classification, enabling precise spatial mapping of cellular distributions.}
\label{fig:annotation_workflow}
\end{figure}

The annotation process consisted of a two-stage protocol (Figure \ref{fig:annotation_workflow}). Initially, a prototype AI model performed automated cell detection and placed preliminary annotation markers across the tissue sections. Subsequently, expert annotators systematically reviewed these AI-generated annotations using the Cu-Cyto\textsuperscript{\tiny{\textregistered}} Viewer interface, performing two primary tasks: (1) verification of AI-placed markers for accuracy, (2) addition of markers for cells missed by the automated detection system.

The annotation system employed a color-coded marking scheme where distinct colors represented different cellular populations and morphological categories. This visual encoding facilitated rapid identification of annotation patterns and enabled efficient quality control assessment across multiple annotators.
Annotation quality was maintained through a multi-stage verification protocol. Following the initial annotation refinement by expert annotators, a second expert annotator performed a double-check of the annotations, with a final review conducted by a pathologist to ensure the highest level of accuracy and reliability. Inter-annotator agreement was assessed through systematic comparison of overlapping annotation regions, ensuring consistent application of classification criteria across all tissue sections.
The Cu-Cyto\textsuperscript{\tiny{\textregistered}} Viewer platform recorded detailed annotation metadata including marker coordinates, classification confidence levels, and annotator identification for traceability. This comprehensive documentation enabled retrospective quality assessment and facilitated standardized patch extraction procedures for subsequent model training and validation processes.

This protocol required complete cellular coverage, ensuring that every identifiable cell within the tissue section received appropriate classification markers, which serves as the basis for calculating the estimated true prior probabilities for cell classification.

\subsection{Dataset Preparation and Sampling Strategy}

To address class imbalance inherent in the original dataset and create balanced training sets suitable for deep learning, we implemented a systematic data selection and augmentation pipeline. The process consisted of three main stages: class-balanced sampling, data augmentation through geometric transformations, and systematic analysis of sampling effects.

\subsubsection{Class-Balanced Sampling}

Data selection was performed using a custom Python implementation that applies stratified sampling to achieve uniform class distribution. The sampling strategy was controlled by two key parameters: \texttt{FlagLimit}, which defines the maximum number of samples per class, and \texttt{FlagMin}, set to 5\% of \texttt{FlagLimit}, which defines the minimum threshold for class inclusion.

\begin{figure}[htbp]
\centering
\includegraphics[width=0.8\columnwidth]{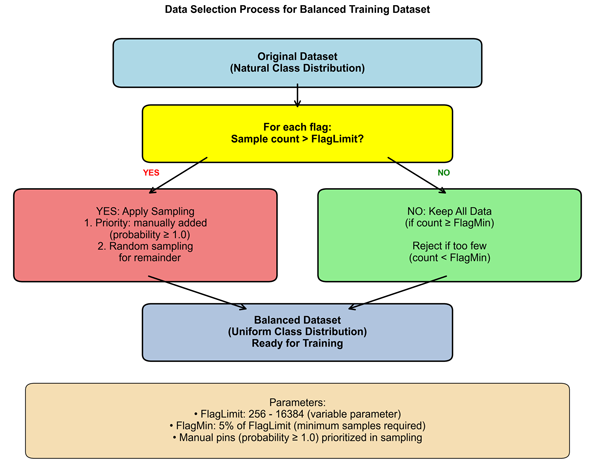}
\caption{Data selection process flowchart for balanced training dataset creation. The decision tree illustrates the transformation from natural class distribution to uniform class distribution through stratified sampling. For each flag class, samples exceeding the FlagLimit threshold undergo priority-based sampling (manually curated samples first, followed by random selection), while classes with insufficient samples are either retained completely or excluded based on the FlagMin threshold. Key parameters include FlagLimit (256-16384), FlagMin (5\% of FlagLimit), and prioritization of manual annotations (probability $\geq$ 1.0).}
\label{fig:selection_flowchart}
\end{figure}

For each class (flag), the selection process followed a hierarchical strategy:

\begin{enumerate}
    \item \textbf{Priority Selection}: Manually curated samples (probability $=$ 1.0) were prioritized for inclusion. These represent cases that were initially misclassified by automated methods but subsequently corrected through expert annotation.
    
    \item \textbf{Random Sampling}: When the total number of samples exceeded \texttt{FlagLimit}, remaining slots were filled through random sampling from the available pool.
    
    \item \textbf{Inclusion Criteria}: Classes with fewer than \texttt{FlagMin} samples were excluded from training to ensure statistical reliability.
\end{enumerate}

Seven balanced datasets were generated using \texttt{FlagLimit} values ranging from 256 to 16,384 to evaluate the effect of dataset size on model performance. This approach transforms the natural class distribution (reflecting real-world prevalence) into uniform distributions suitable for supervised learning.

\subsubsection{Sampling Analysis and Validation}

The effectiveness of the sampling strategy was evaluated through systematic analysis of the selection process across all seven \texttt{FlagLimit} conditions (256, 512, 1024, 2048, 4096, 8192, and 16384). From an original dataset of 185,432 samples across 20 unique flags, the sampling process achieved substantial data reduction while maintaining class balance.

The sampling process demonstrated consistent performance across different \texttt{FlagLimit} values. Overall data reduction rates ranged from 97.7\% at the most restrictive setting (FlagLimit = 256) to 37.7\% at the most permissive setting (FlagLimit = 16384). Average usage rates showed a steady increase from 32.0\% to 86.7\% as \texttt{FlagLimit} increased, following a characteristic curve with steeper initial growth that gradually levels off at higher limits.

\begin{figure}[htbp]
\centering
\includegraphics[width=0.9\columnwidth]{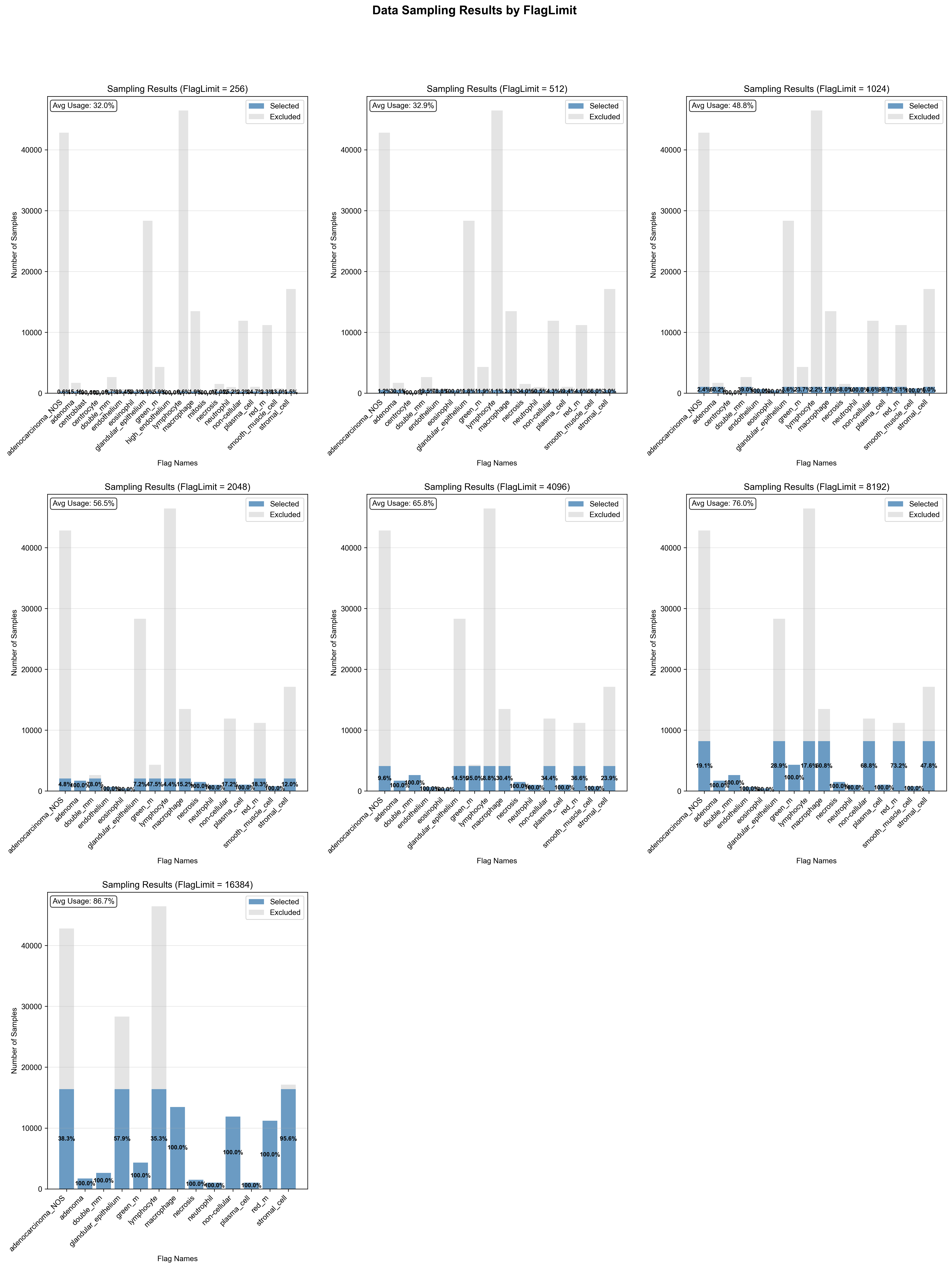}
\caption{Sampling results across different FlagLimit values. Each panel shows stacked bar charts displaying selected (blue) and excluded (gray) samples for each flag class. Usage rates are indicated as percentages on the bars. The progression from FlagLimit = 256 to FlagLimit = 16384 demonstrates increasing utilization of available samples while maintaining class balance. Note that some flags are excluded at higher FlagLimit values due to insufficient samples meeting the FlagMin threshold.}
\label{fig:sampling_results}
\end{figure}

The number of processed flags varied with \texttt{FlagLimit} due to the \texttt{FlagMin} threshold requirement. At restrictive settings (FlagLimit = 256), 20 flags met the inclusion criteria, while at higher settings, some rare classes were excluded, resulting in 13-17 flags being processed. This demonstrates the trade-off between dataset size and class diversity inherent in balanced sampling strategies.

Analysis of individual flag usage patterns revealed significant heterogeneity in sample availability and selection rates across different classes. Usage rate heatmaps demonstrated that certain flags (e.g., smooth muscle cell, neutrophil, centroblast) achieved near-complete utilization even at moderate \texttt{FlagLimit} values, indicating abundant original samples. Conversely, rarer classes such as stromal\_cell and non-cellular required higher \texttt{FlagLimit} values to achieve substantial usage rates, reflecting their limited representation in the original dataset.

The proportion of flags achieving full usage ($\geq$ 99.9\%) increased systematically with \texttt{FlagLimit}: from 4/20 flags at the most restrictive setting to 9/13 flags at the most permissive setting. This progression illustrates the sampling algorithm's effectiveness in preferentially utilizing available samples while maintaining the target class balance.

\begin{figure}[htbp]
\centering
\includegraphics[width=\columnwidth]{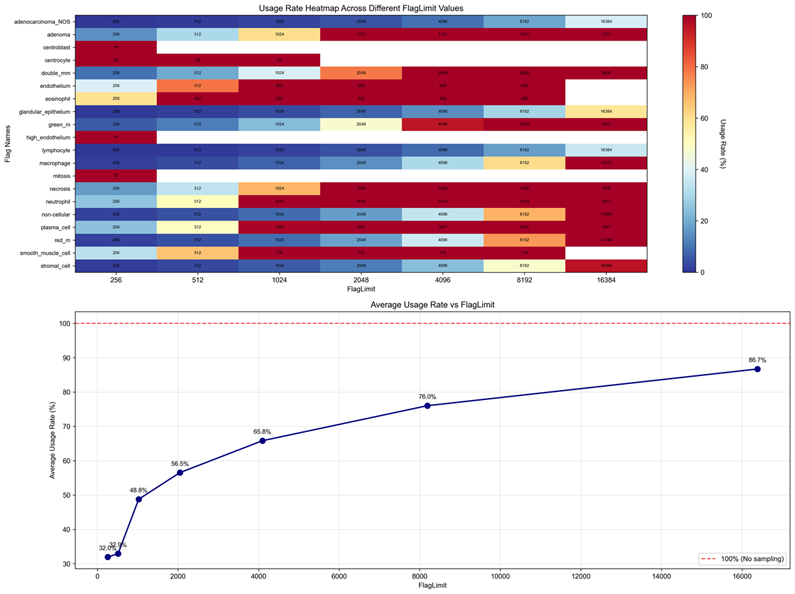}
\caption{Usage rate analysis across FlagLimit values. Top: Heatmap showing usage rates (\%) for each flag class across different FlagLimit settings. Color intensity represents usage rate from 0\% (blue) to 100\% (red). Bottom: Average usage rate progression as a function of FlagLimit, showing the characteristic curve with steeper initial growth that gradually levels off. The horizontal dashed line at 100\% represents complete sample utilization.}
\label{fig:usage_analysis}
\end{figure}

\section{Methods}

\subsection{ResNet Architecture Implementation}

The ResNet model was implemented using a four-stage residual block architecture with systematic channel progression. The network employed a block distribution strategy where the total number of residual blocks $n_{block}$ was partitioned according to $4a + b = n_{block}$, with $a = \lfloor n_{block}/4 \rfloor$ and $b = n_{block} \bmod 4$. This distribution allocated blocks across four stages as follows: $n_{block1} = a$, $n_{block2} = a$, $n_{block3} = a + b$, and $n_{block4} = a$, with corresponding channel dimensions of 24, 48, 64, and 128, respectively.

The architecture initiated with a 3×3 convolutional layer transforming input channels to 24 feature maps, followed by four sequential residual block stages. Each residual block consisted of two 3×3 convolutional layers with ReLU activation and skip connections. 
A distinctive feature of the first stage was the incorporation of RGB channel concatenation, where original RGB channels were systematically concatenated with processed feature maps at three sequential points within the ResNet final processing layers, preserving low-level color information throughout the high-level feature extraction process. Spatial downsampling was achieved through 2×2 max pooling operations between stages, reducing the spatial resolution from 40×40 to 5×5 pixels across the four stages.

The final portion of the network comprised three fully connected layers with dimensions 3200$\rightarrow$512$\rightarrow$256$\rightarrow$$n_{classes}$, where 3200 corresponds to the flattened feature map size (128 × 5 × 5). Dropout regularization was applied with ratios of 0.1 for input layers, 0.5 for hidden layers, and 0.0 for convolutional layers. The model was optimized using stochastic gradient descent with momentum 0.5, L2 weight decay of 5×10$^{-4}$, and a learning rate scaled by a factor of 6 relative to the baseline configuration to accelerate convergence.

\subsubsection{Prior Bias Quantification}

To quantify the magnitude of prior distribution shifts between training and target populations, we defined a bidirectional bias metric that captures both overrepresentation and underrepresentation scenarios. For each class $c$, the bias ratio is calculated as:

\begin{equation}
\text{bias\_ratio}_c = \frac{P_{\text{true}}(Y = c)}{P_{\text{training}}(Y = c)}
\end{equation}

where $P_{\text{true}}(Y = c)$ represents the true prior probability in the target population and $P_{\text{training}}(Y = c)$ represents the prior probability in the training dataset.

To account for both directions of bias (overrepresentation when bias\_ratio $> 1$ and underrepresentation when bias\_ratio $< 1$), we compute the inverse bias ratio:

\begin{equation}
\text{inverse\_bias\_ratio}_c = \frac{1}{\text{bias\_ratio}_c} = \frac{P_{\text{training}}(Y = c)}{P_{\text{true}}(Y = c)}
\end{equation}

The maximum bias for each class is defined as the larger absolute deviation from unity:

\begin{align}
&\text{max\_bias\_for\_class}_c \nonumber \\ &= \max(|\text{bias\_ratio}_c|, |\text{inverse\_bias\_ratio}_c|)
\end{align}

This bidirectional approach ensures that both scenarios—where a class is overrepresented in training data relative to the target population (high bias\_ratio) and where it is underrepresented (high inverse\_bias\_ratio)—are appropriately captured. For example, if a class has a true prior of 0.1 but training prior of 0.2, the bias\_ratio would be 0.5, while the inverse\_bias\_ratio would be 2.0, resulting in a max\_bias\_for\_class of 2.0.

\begin{figure}[htbp]
\centering
\includegraphics[width=\columnwidth]{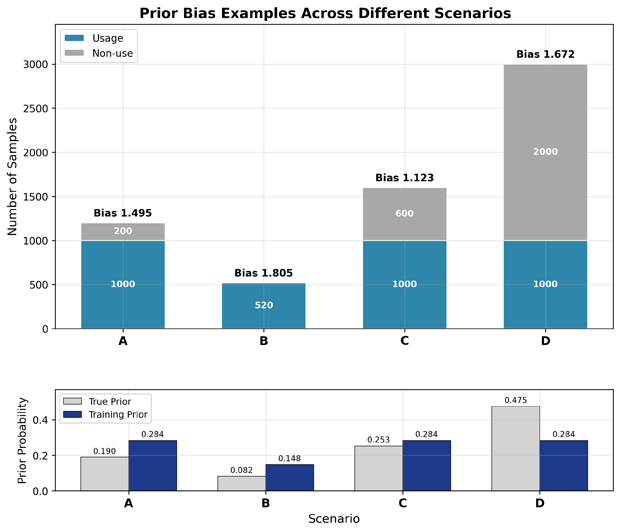}
\caption{Examples of prior bias scenarios across different cell types. Top panel shows sample composition with usage (blue) and non-use (gray) proportions for each scenario. Bottom panel compares true prior probabilities (light gray) and training prior probabilities (dark blue) side by side. The bias values shown above each scenario (A: 1.495, B: 1.805, C: 1.123, D: 1.672) represent the maximum bias calculated using the bidirectional metric, capturing both overrepresentation and underrepresentation relative to true population priors.}
\label{fig:prior_bias_examples}
\end{figure}

\subsection{Similarity-based Separability Analysis}

To evaluate the quality of learned feature representations, we assess how well the feature space separates different classes by analyzing the similarity patterns between test samples and training samples.

For a test sample $x_{\text{test}} \in C_i$ of class $C_i$, we compute the similarity with training samples from the same class $\mathcal{T}_i = \{x_1^{(i)}, x_2^{(i)}, ..., x_{n_i}^{(i)}\}$:

\begin{equation}
S_{\text{intra}}(x_{\text{test}}, C_i) = \frac{1}{k} \sum_{j \in \text{Top-k}} \text{cos\_sim}(\phi(x_{\text{test}}), \phi(x_j^{(i)}))
\end{equation}

where Top-k represents the $k$ samples with the highest similarity scores. This metric captures how closely the test sample aligns with its own class in the feature space.

For the same test sample $x_{\text{test}} \in C_i$, we compute the similarity with training samples from other classes $C_j (j \neq i)$:

\begin{align}
&S_{\text{inter}}(x_{\text{test}}, C_{\neg i}) \nonumber \\ &= \frac{1}{k \cdot (K-1)} \sum_{j \neq i} \sum_{l \in \text{Top-k}} \text{cos\_sim}(\phi(x_{\text{test}}), \phi(x_l^{(j)}))
\end{align}

This measures the average similarity between the test sample and the most similar samples from all other classes, providing insight into potential class confusion in the feature space.

We define the separability score as the difference between intra-class and inter-class similarities:

\begin{equation}
\Delta S(x_{\text{test}}) = S_{\text{intra}}(x_{\text{test}}, C_i) - S_{\text{inter}}(x_{\text{test}}, C_{\neg i})
\end{equation}

A positive $\Delta S$ indicates that the test sample is more similar to its own class than to other classes, suggesting good feature space separability. Based on an empirically determined threshold $\tau$, we classify each sample's separability quality:

\begin{equation}
\text{Separability Quality} = \begin{cases}
\text{Good} & \text{if } \Delta S(x_{\text{test}}) \geq \tau \\
\text{Poor} & \text{if } \Delta S(x_{\text{test}}) < \tau
\end{cases}
\end{equation}

The threshold $\tau$ is determined empirically through accuracy analysis to identify samples for which the learned feature representation provides reliable class discrimination.

\subsection{Margin-based Confidence Analysis}

The learned features $\Phi_\theta(\mathbf{x})$ are mapped to class logits through the final classification layer:
\begin{equation}
\mathbf{z} = W^T \Phi_\theta(\mathbf{x}) + \mathbf{b}
\end{equation}

where $W$ and $\mathbf{b}$ are the weight matrix and bias vector of the classification layer. The posterior probability for each class is then computed using the softmax function:
\begin{equation}
P_\theta(Y = c | \mathbf{x}) = \text{softmax}(\mathbf{z})_c = \frac{e^{z_c}}{\sum_{j=1}^{K} e^{z_j}}
\end{equation}

For each test sample, we compute the prediction margin:
\begin{equation}
M_c(\mathbf{x}) = P_\theta(Y = c | \mathbf{x}) - \max_{c' \neq c} P_\theta(Y = c' | \mathbf{x})
\end{equation}

This margin serves as an indicator of the posterior probability's reliability, where higher margins suggest more confident predictions based on better feature separability.

According to Bayes' theorem, the posterior probability is given by:
\begin{equation}
P(Y = c | \mathbf{x}) = \frac{P(\Phi_\theta(\mathbf{x})|C_c) \cdot P(Y = c)}{\sum_{j=1}^{K} P(\Phi_\theta(\mathbf{x})|C_j) \cdot P(Y = j)}
\end{equation}

While the previous similarity-based analysis evaluates the quality of learned likelihood estimation $P(\Phi(x)|C)$ in the feature space, margin analysis assesses the reliability of the final posterior probability $P(Y=c|x)$ output by the model. This distinction is crucial because the posterior probability depends not only on likelihood quality but also on the classification layer and prior probability handling.

\subsection{Statistical Analysis}

\subsubsection{Correlation Analysis}

We first examine the bivariate relationships between F1 score or margin scores and potential predictor variables using Pearson correlation coefficients.
%For each numerical variable in the dataset, we calculate the correlation with the margin score, which represents the prediction confidence:
%\begin{equation}
%r = \frac{\sum_{i=1}^{n}(x_i - \bar{x})(y_i - \bar{y})}{\sqrt{\sum_{i=1}^{n}(x_i - \bar{x})^2\sum_{i=1}^{n}(y_i - \bar{y})^2}}
%\end{equation}
Statistical significance is assessed using p-values, with significance levels marked as: *** (p $<$ 0.001), ** (p $<$ 0.01), * (p $<$ 0.05).

\subsubsection{Linear Regression Analysis}
We perform multiple linear regression to quantify the relationships between margin scores and our key predictor variables:

\begin{align}
\text{F1\_score} &= \beta_0 + \beta_1 \cdot \text{lq\_quality\_score} \nonumber \\ &+ \beta_2 \cdot \text{bias\_max\_for\_class} \nonumber \\ &+ \beta_3 \cdot \text{sa\_adequacy\_ratio}+ \epsilon
\end{align}

where `lq\_quality\_score` represents the likelihood quality (feature separability) ,`bias\_max\_for\_class` represents the prior bias level and `sa\_adequacy\_ratio` represents the sample adequacy ratio. Model diagnostics and coefficient significance are evaluated using ordinary least squares (OLS) regression.

\subsubsection{Analysis of Variance (ANOVA)}

\textbf{One-way ANOVA:}
We conduct separate one-way ANOVA tests to assess the individual effects of feature separability and prior bias on margin distribution.

%\begin{equation}
%H_0: \mu_1 = \mu_2 = ... = \mu_k \quad \text{vs} \quad H_1: \text{At least one } \mu_i \neq \mu_j
%\end{equation}
%where $\mu_i$ represents the population mean of the dependent variable (e.g., margin, F1 score) for group $i$.

For continuous variables, we create categorical groups using quartile-based binning to enable ANOVA analysis. Effect sizes are quantified using eta-squared ($\eta^2$):

\begin{equation}
\eta^2 = \frac{SS_{\text{between}}}{SS_{\text{total}}}
\end{equation}

\textbf{Two-way ANOVA (Multiple Linear Regression Framework):} 
We perform analysis of variance using a linear model framework to examine both main effects and interaction effects between continuous predictors.

%\begin{align}
%M &= \beta_0 + \beta_1 \cdot \text{LQ\_Score} + \beta_2 \cdot \text{Bias\_Level} \nonumber \\ &+ \beta_3 \cdot (\text{LQ\_Score} \times \text{Bias\_Level}) + \epsilon
%\end{align}

%where:
%\begin{itemize}
%\item $\beta_1$: Main effect of likelihood quality score (continuous)
%\item $\beta_2$: Main effect of prior bias level (continuous)  
%\item $\beta_3$: Interaction effect between likelihood quality and bias level
%\item $\epsilon$: Random error term
%\end{itemize}

%The ANOVA framework decomposes the total variance into components attributable to each main effect and their interaction, allowing us to test:
%\begin{align}
%H_0^{(1)}: & \beta_1 = 0 \ \text{(No main effect of feature separability)} \\
%H_0^{(2)}: & \beta_2 = 0 \ \text{(No main effect of prior bias)} \\
%H_0^{(3)}: & \beta_3 = 0 \ \text{(No interaction effect)}
%\end{align}

%\textbf{Hypothesis Testing:}
%Our primary hypothesis predicts that feature separability has a significantly stronger effect on margin distribution than prior bias:
%\begin{align}
%H_1: & \text{Effect of likelihood quality} \nonumber \\ &\quad >> \text{Effect of prior bias} \\
%H_2: & \text{Interaction effect} \approx 0
%\end{align}

All statistical analyses are performed using Python with scipy.stats for correlation and ANOVA tests, and statsmodels for regression analysis. Results include F-statistics, p-values, and effect sizes with appropriate visualization through box plots and interaction plots.

\subsubsection{Generalized Additive Model (GAM) Analysis}

To capture potential non-linear relationships and complex interactions between predictors that linear models may miss, we employed Generalized Additive Models (GAM). GAMs extend linear regression by replacing linear terms with smooth functions, allowing for flexible modeling of non-linear relationships while maintaining interpretability.

\textbf{Model Evaluation:}
GAM performance was assessed using:
\begin{itemize}
\item $R^2$: Proportion of variance explained
\item AIC (Akaike Information Criterion): Model fit penalized for complexity
\item Effective degrees of freedom: Measure of model complexity
\item Residual diagnostics: Assessment of model assumptions
\end{itemize}

\textbf{Implementation:}
GAMs were fitted using Python's pygam library. Statistical significance of smooth terms was evaluated using approximate p-values. Visual interpretation of partial dependence plots and interaction surfaces was prioritized alongside statistical measures for model understanding.

\textbf{Comparative Analysis:}
We fitted parallel GAM models using both F1 scores and prediction margins as dependent variables to examine potential differences in optimal operating conditions between classification performance and model confidence, providing insights into confidence calibration issues.

\section{Results}

\subsection{ResNet Performance}

\begin{figure}[htbp]
\centering
\includegraphics[width=\columnwidth]{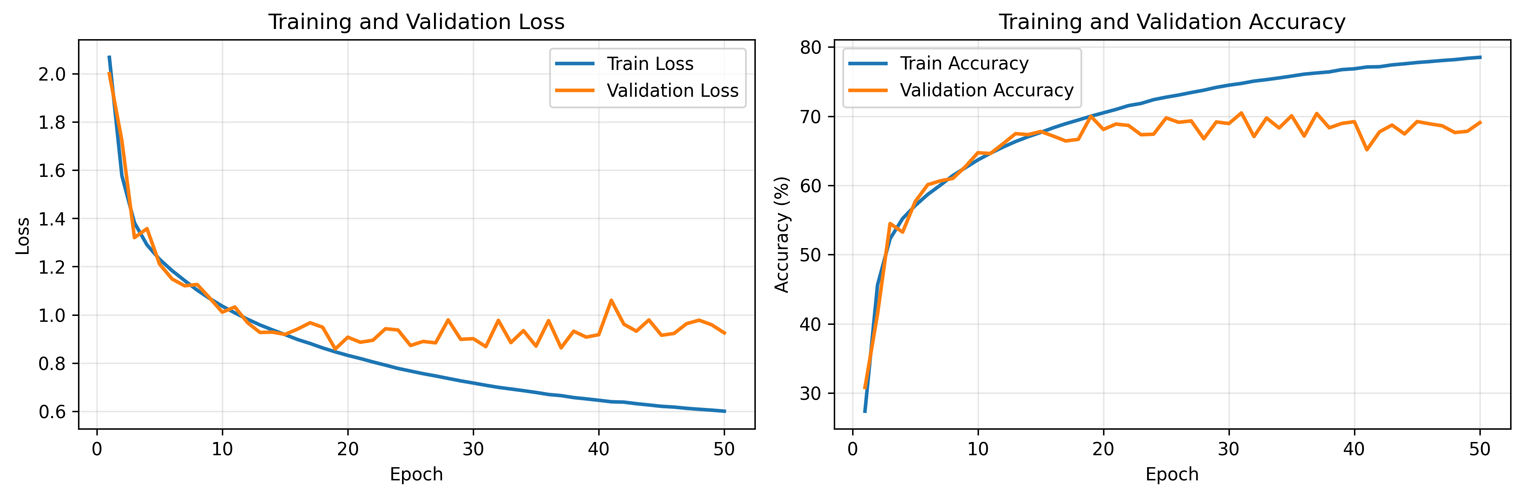}
\caption{ResNet training dynamics showing loss and accuracy trajectories over 50 epochs. Left panel displays training and validation loss curves demonstrating exponential decay with early stabilization. Right panel shows corresponding accuracy improvements with plateau behavior after epoch 20, indicating convergence of the optimization process.}
\label{fig:resnet_training}
\end{figure}

Training of the ResNet model under FlagLimit 8192 conditions demonstrated stable convergence characteristics over 50 epochs (Figure \ref{fig:resnet_training}). The model achieved a final training accuracy of 77\% and validation accuracy of 69\%, indicating moderate overfitting with an 8\% generalization gap. Loss trajectories showed exponential decay patterns, with training loss decreasing from 2.0 to 0.6 and validation loss stabilizing at approximately 0.9 after initial rapid decline. The learning dynamics exhibited rapid initial improvement during the first 10 epochs, followed by gradual refinement. Validation accuracy plateaued around epoch 20, with subsequent training showing minimal improvement in generalization performance.

\begin{figure}[htbp]
\centering
\includegraphics[width=0.8\columnwidth]{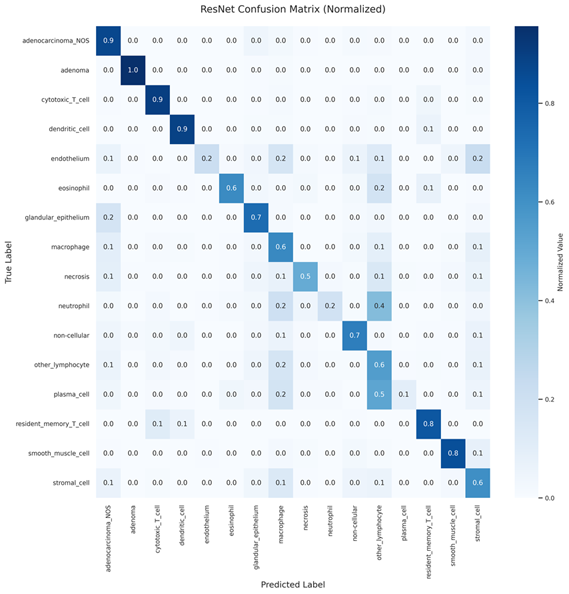}
\caption{Normalized confusion matrix for ResNet classification across 16 cell types. Diagonal elements represent correct classification rates, while off-diagonal elements indicate misclassification patterns. High-performing classes (adenocarcinoma NOS, adenoma, cytotoxic T cell, dendritic cell) show strong diagonal signals, whereas challenging classes (endothelium, neutrophil, plasma cell) exhibit substantial confusion with morphologically similar cell types.}
\label{fig:resnet_confusion}
\end{figure}

Classification performance varied substantially across cell types, as evidenced by the confusion matrix analysis (Figure \ref{fig:resnet_confusion}). High-performance classes included adenocarcinoma NOS (90\% accuracy), adenoma (100\% accuracy), cytotoxic T cell (90\% accuracy), and dendritic cell (90\% accuracy). Conversely, several cell types exhibited poor classification performance: endothelium (20\% accuracy), neutrophil (20\% accuracy), and plasma cell (10\% accuracy). The overall system accuracy reached 70\% across all 16 cell types.

Misclassification patterns revealed systematic confusion between morphologically similar cell types. The most frequent errors occurred between macrophage and other lymphocyte classes, and between various stromal cell populations. These patterns suggest that current feature representations may be insufficient to capture subtle morphological distinctions between closely related cell types.

\begin{figure}[htbp]
\centering
\includegraphics[width=0.9\columnwidth]{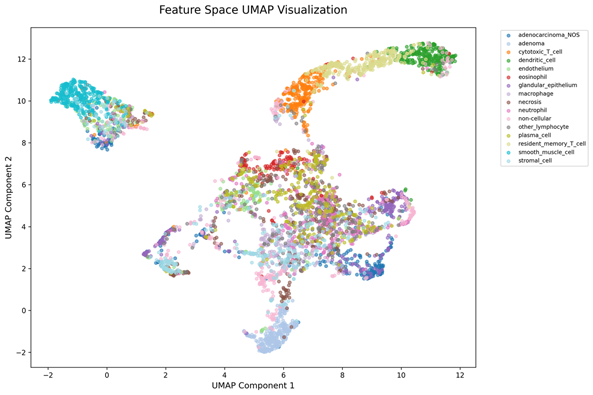}
\caption{UMAP visualization of ResNet learned feature representations for test samples. Each point represents a single cell image projected into 2D feature space, colored by true cell type. Well-separated clusters (adenocarcinoma NOS, adenoma, cytotoxic T cell) correspond to high classification accuracy, while overlapping regions indicate challenging discrimination tasks. The feature space topology directly correlates with confusion matrix patterns, demonstrating the model's learned hierarchical relationships between cell types.}
\label{fig:resnet_umap}
\end{figure}

Feature space analysis through UMAP dimensionality reduction revealed the learned representation structure of the ResNet model (Figure \ref{fig:resnet_umap}). Well-separated clusters emerged for high-performing cell types, including distinct regions for adenocarcinoma NOS, adenoma, and cytotoxic T cell populations. Conversely, poorly classified cell types showed substantial overlap in the reduced feature space, particularly among stromal cell populations and various lymphocyte subtypes.

\subsection{Likelihood Quality Analysis}

\begin{figure}[htbp]
\centering
\includegraphics[width=\columnwidth]{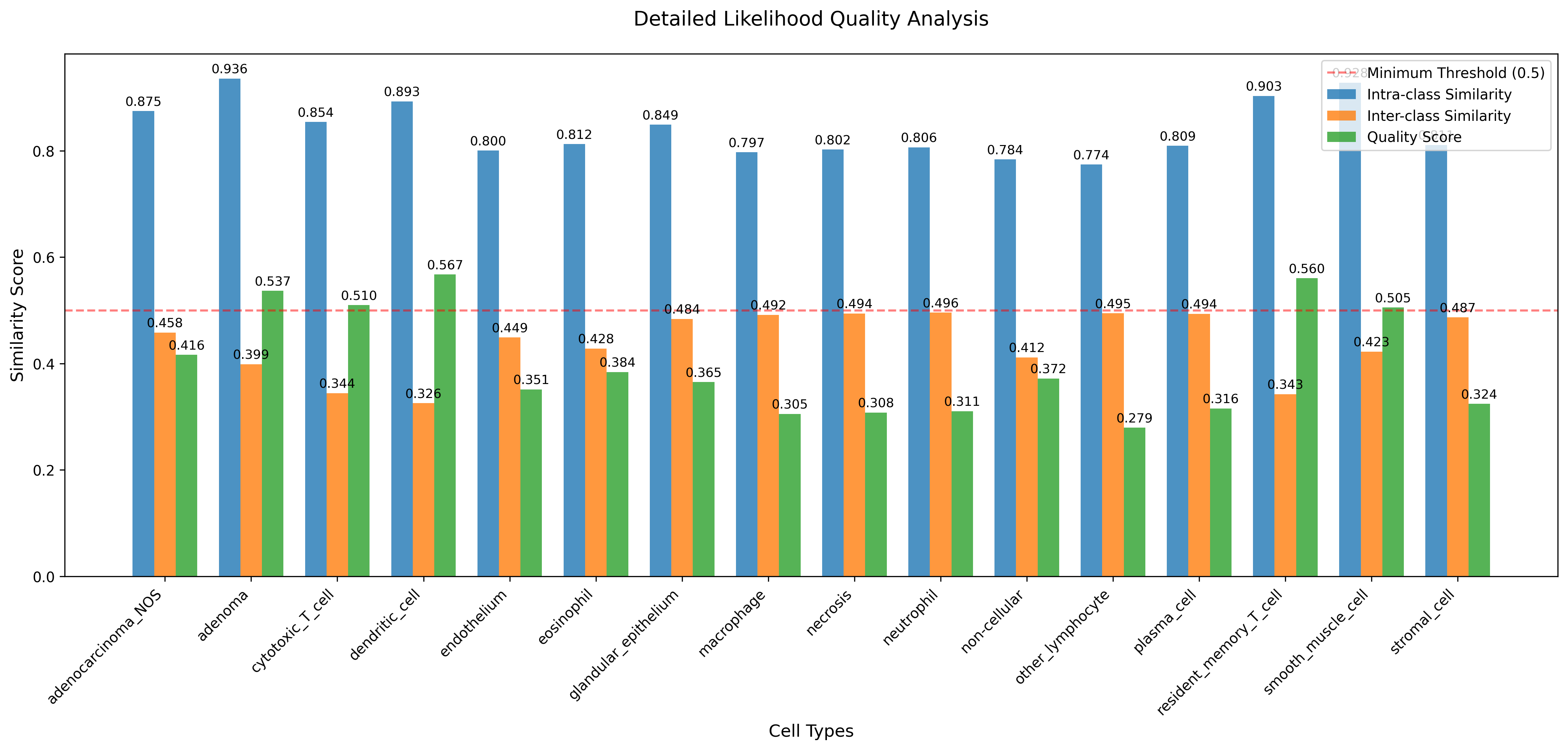}
\caption{Detailed likelihood quality analysis across 16 cell types. Blue bars represent intra-class similarity (higher values indicate better within-class coherence), orange bars show inter-class similarity (lower values indicate better between-class separation), and green bars display the resulting quality scores (intra-class minus inter-class similarity). The red dashed line indicates the minimum quality threshold of 0.5 for reliable classification performance.}
\label{fig:likelihood_quality}
\end{figure}

To quantitatively assess the relationship between feature separability and classification performance, we conducted a detailed likelihood quality analysis across all 16 cell types (Figure \ref{fig:likelihood_quality}). This analysis examines the geometric relationships between intra-class and inter-class similarity measures in the learned feature space.

The likelihood quality scores revealed a clear correspondence with classification performance observed in the confusion matrix analysis. High-performing cell types demonstrated superior feature separability characteristics: adenoma achieved the highest quality score (0.558), followed by dendritic cell (0.556) and resident memory T cell (0.577). These scores reflect strong intra-class similarity (0.937, 0.894, and 0.925 respectively) combined with relatively low inter-class similarity (0.380, 0.338, and 0.326 respectively).

Conversely, poorly performing cell types exhibited quality scores substantially below the 0.3 threshold. Most notably, other lymphocyte demonstrated the lowest quality score (0.280), corresponding to its poor classification performance (confusion matrix accuracy below 60\%). Similarly, macrophage (0.310), necrosis (0.308), and neutrophil (0.310) clustered around the quality threshold, aligning with their observed classification difficulties.

\subsection{Decision Boundary Margin Analysis}
\begin{figure}[htbp]
\centering
\includegraphics[width=\columnwidth]{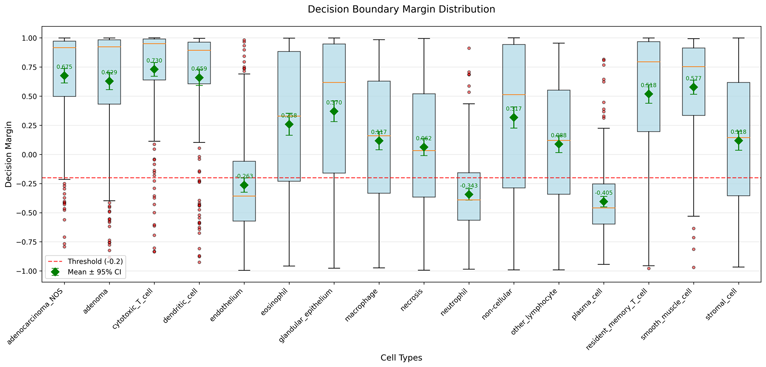}
\caption{Decision boundary margin distribution across 16 cell types. Box plots show the distribution of prediction margins (true label probability minus highest probability among other classes) for each cell type. Green diamonds represent mean values with 95\% confidence intervals. The red dashed line indicates the margin threshold of -0.2 for uncertain predictions. High-performing cell types demonstrate positive margins with narrow distributions, while problematic cell types show negative or highly variable margins.}
\label{fig:margin_distribution}
\end{figure}

The decision boundary margin analysis examines the distribution of decision margins—defined as the difference between the predicted probability of the true label and the highest probability among all other classes—across all 16 cell types (Figure \ref{fig:margin_distribution}). This analysis provides insights into model confidence patterns across different cellular populations.

The margin analysis revealed substantial disparities between cell types. High-performing cell types demonstrated consistently positive margins: adenocarcinoma NOS (mean = 0.675), adenoma (mean = 0.629), cytotoxic T cell (mean = 0.730), and dendritic cell (mean = 0.659). These populations exhibited narrow confidence intervals and relatively compact distributions.

Conversely, poorly performing cell types showed strongly negative margins. Endothelium demonstrated negative margins (mean = -0.263), while neutrophil (mean = -0.343) and plasma cell (mean = -0.405) exhibited the most negative margin patterns among all cell types.

The margin threshold analysis at -0.2 (indicated by the red dashed line) effectively separated well-performing from poorly-performing cell types. Cell types falling below this threshold included endothelium, neutrophil, and plasma cell, with other lymphocyte hovering near the boundary (mean = 0.088).

\subsection{Multiple Linear Regression Analysis}

\begin{figure}[htbp]
\centering
\includegraphics[width=0.9\columnwidth]{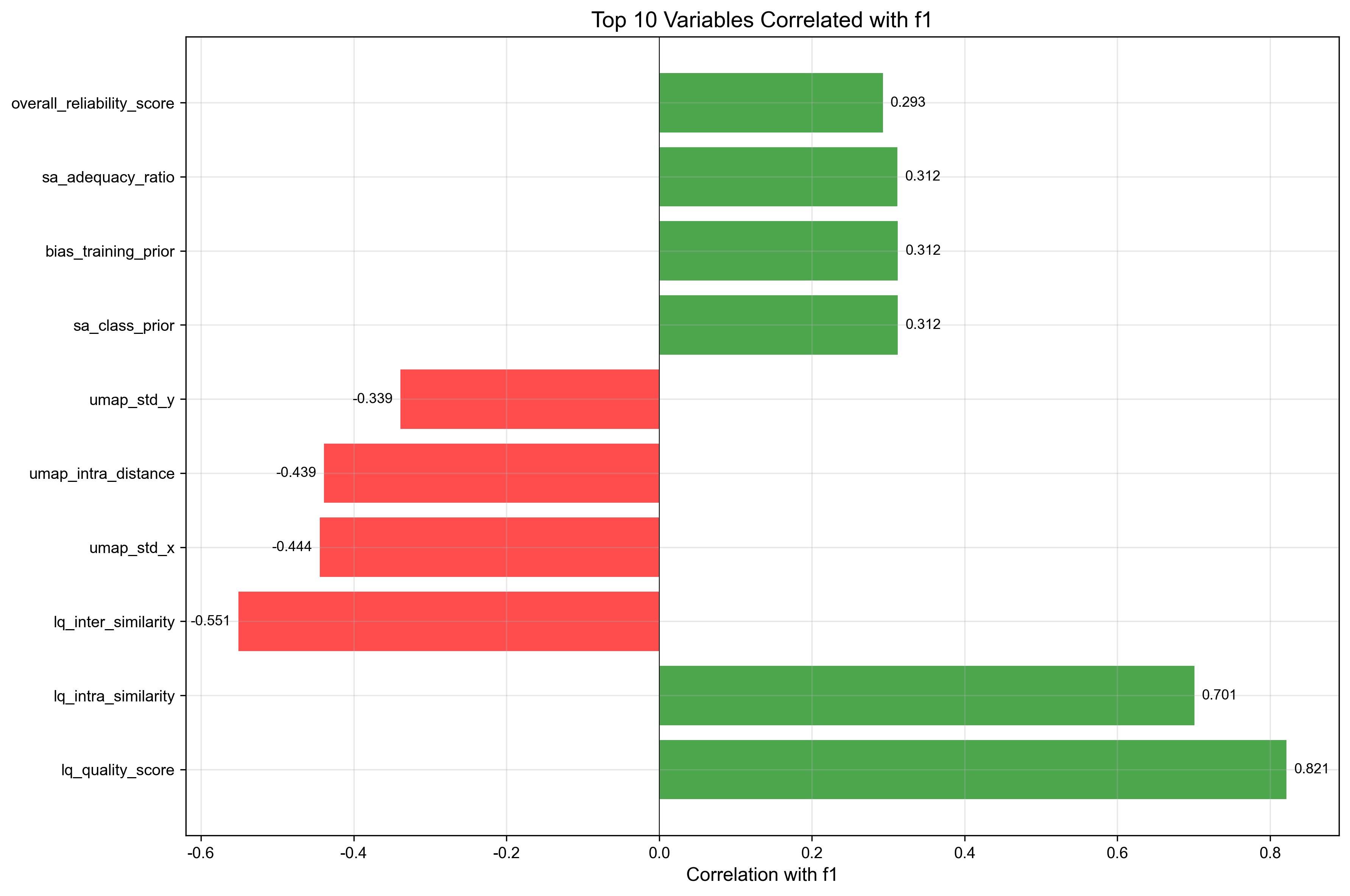}
\caption{Top 10 variables correlated with F1 scores. Green bars indicate positive correlations, red bars indicate negative correlations. The three variables used in regression analysis include the strongest predictor (likelihood quality score) and two moderate predictors.}
\label{fig:correlation_plot}
\end{figure}

Correlation analysis identified the strongest predictors of F1 performance after excluding redundant features within variable categories (Figure \ref{fig:correlation_plot}).

\begin{figure}[htbp]
\centering
\includegraphics[width=0.9\columnwidth]{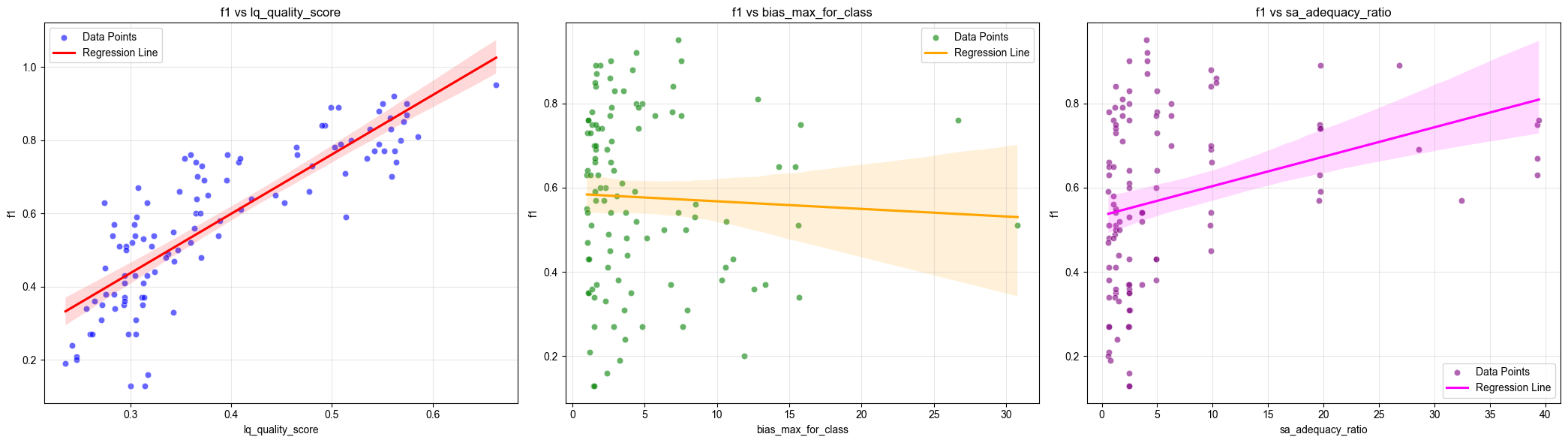}
\caption{Regression analysis of F1 scores against three key predictors. Left panel shows the strong positive relationship between likelihood quality score and F1 scores. Middle panel demonstrates a weak negative relationship between bias maximum for class and F1 scores. Right panel illustrates the positive relationship between sample adequacy ratio and F1 scores.}
\label{fig:regression_plots}
\end{figure}

Among the independent variables analyzed, likelihood quality score showed the strongest correlation ($r = 0.821$), followed by likelihood intra-similarity ($r = 0.701$). However, to capture distinct aspects of model performance while avoiding multicollinearity, three predictors were selected from different variable categories: likelihood quality score (representing embedding quality, $r = 0.821$), sample adequacy ratio (representing training data sufficiency, $r = 0.312$), and bias max for class (representing class imbalance effects, $r = -0.044$)(Figure \ref{fig:regression_plots}). This selection strategy ensures that the regression model incorporates complementary information from quality metrics, sample characteristics, and bias patterns, rather than redundant information from highly correlated variables within the same category.

Multiple linear regression analysis revealed that all three predictors are significant determinants of F1 performance. The overall model achieved exceptional explanatory power ($R^2 = 0.826$, $F(3, 105) = 166.0$, $p < 0.001$), accounting for approximately 83\% of the variance in F1 scores. The fitted regression model is expressed as:

\begin{align}
&\text{F1 Score} = -0.132 + 1.650 \times \text{LQ Score} \nonumber \\ &+ 0.004 \times \text{Bias Max} + 0.008 \times \text{Adequacy Ratio}
\label{eq:regression_model}
\end{align}

\begin{table}[htbp]
\centering
\caption{Multiple Linear Regression Results for F1 Score Prediction}
\label{tab:regression_results}
\adjustbox{width=\columnwidth}{		% dcの場合
\begin{tabular}{lcccc}
\hline
Variable & Coefficient & Std. Error & $t$-value & $p$-value \\
\hline
Constant & -0.132 & 0.034 & -3.850 & $< 0.001$ \\
Likelihood Quality Score & 1.650 & 0.079 & 20.967 & $< 0.001$ \\
Bias Max for Class & 0.004 & 0.002 & 2.395 & 0.018 \\
Adequacy Ratio & 0.008 & 0.001 & 8.503 & $< 0.001$ \\
\hline
\multicolumn{5}{l}{$N = 109$, $R^2 = 0.826$, $F(3, 105) = 166.0$, $p < 0.001$} \\
\hline
\end{tabular}
}	% dcの場合
\end{table}
The likelihood quality score emerged as the dominant predictor ($\beta = 1.650$, $t = 20.967$, $p < 0.001$), consistent with its strong bivariate correlation with F1 performance ($r = 0.821$). The adequacy ratio also showed a highly significant positive relationship ($\beta = 0.008$, $t = 8.503$, $p < 0.001$). Bias max for class showed a significant positive relationship ($\beta = 0.004$, $t = 2.395$, $p = 0.018$) in the linear model.

\subsection{ANOVA Analysis}

To examine potential non-linear relationships, we first conducted one-way ANOVA analyses for each predictor after converting continuous variables into quartile-based categories. Subsequently, we performed two-way ANOVA analyses to investigate potential interactions between predictors.

\subsubsection{One-way ANOVA}

\begin{figure}[htbp]
\centering
\begin{subfigure}{0.32\columnwidth}
\centering
\includegraphics[width=\columnwidth]{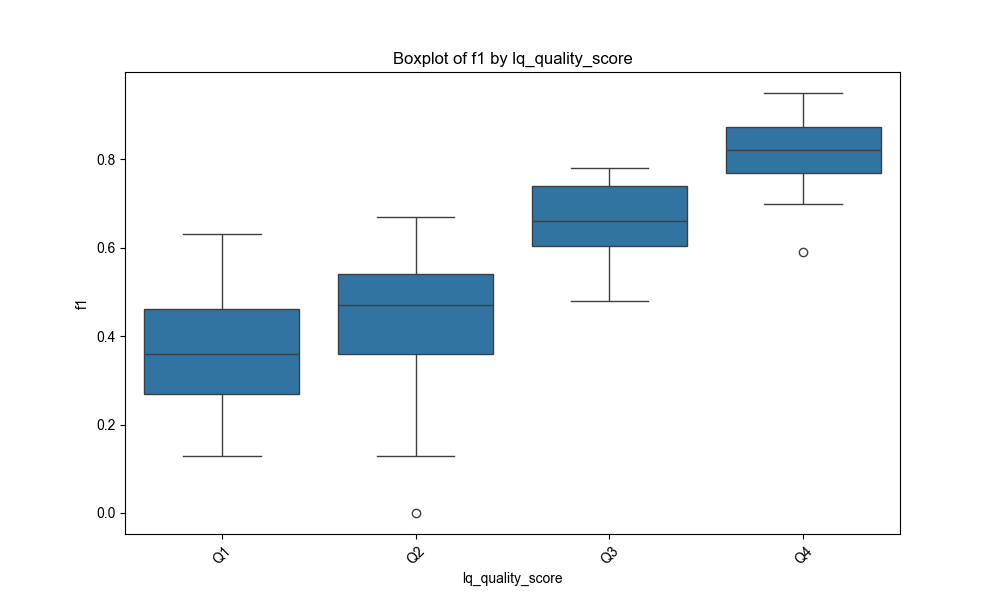}
\caption{Feature Separability}
\end{subfigure}
\hfill
\begin{subfigure}{0.32\columnwidth}
\centering
\includegraphics[width=\columnwidth]{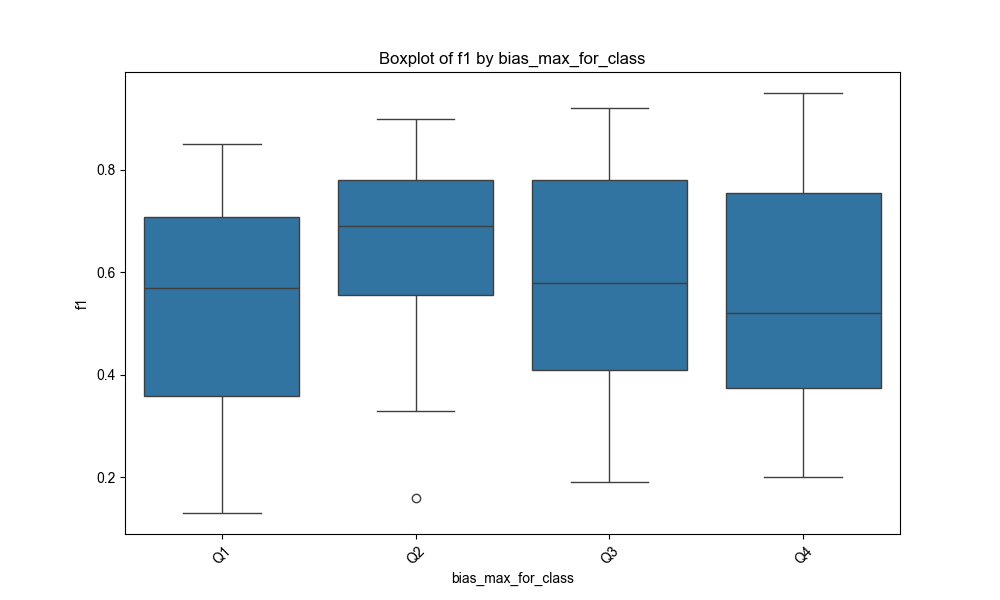}
\caption{Prior Bias}
\end{subfigure}
\begin{subfigure}{0.32\columnwidth}
\centering
\includegraphics[width=\columnwidth]{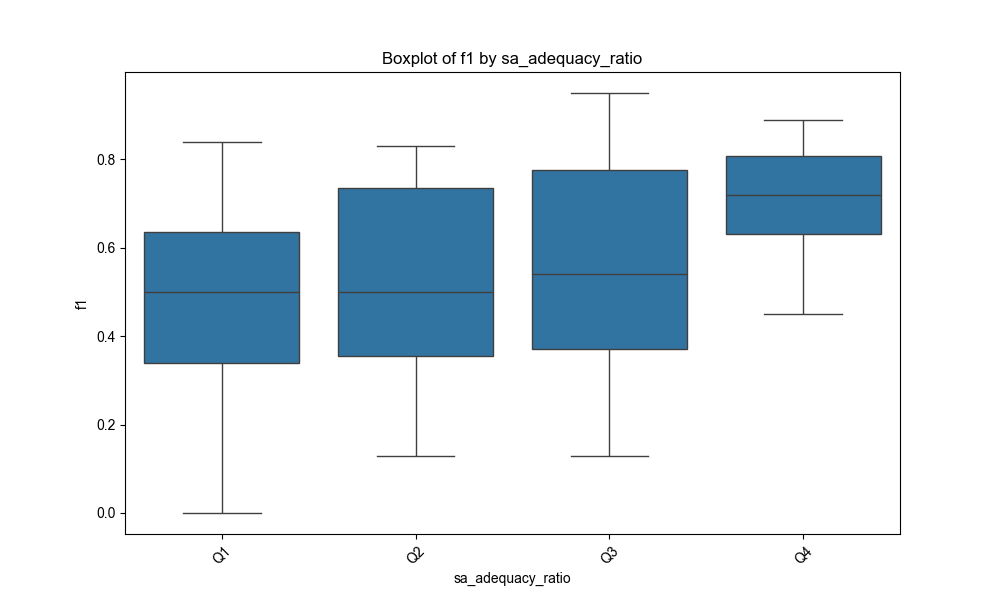}
\caption{Sample Adequacy}
\end{subfigure}
\caption{F1 score distributions across quartiles for each predictor variable.}
\label{fig:boxplots}
\end{figure}

The box plots (Figure \ref{fig:boxplots}) revealed distinct patterns across quartiles: likelihood quality score showed a clear monotonic increase in F1 performance from Q1 to Q4, confirming the strong linear relationship. Adequacy ratio also demonstrated increasing performance with higher quartiles, consistent with the positive regression coefficient. Bias max for class exhibited a non-monotonic pattern across quartiles, with Q2 showing the highest median F1 scores and Q4 showing the lowest, suggesting a more complex relationship with F1 performance.

\subsubsection{Two-way ANOVA for Interaction Effects}

To further investigate potential interaction effects between predictors, we conducted two separate two-way ANOVA analyses examining the interactions between likelihood quality score and the other key predictors. Table \ref{tab:anova1} presents the results for the interaction between likelihood quality score and bias max for class, while Table \ref{tab:anova2} shows the results for likelihood quality score and sample adequacy ratio.

\begin{table}[htbp]
\centering
\caption{Two-way ANOVA results for F1 score with likelihood quality score and bias max for class as factors}
\label{tab:anova1}
\adjustbox{width=\columnwidth}{		% dcの場合
\begin{tabular}{lrrrr}
\hline
\textbf{Source} & \textbf{Sum Sq} & \textbf{df} & \textbf{F} & \textbf{p-value} \\
\hline
LQ Quality Score & 3.264 & 1 & 251.81 & $< 0.001^{***}$ \\
Bias Max for Class & 0.001 & 1 & 0.04 & 0.833 \\
LQ Score $\times$ Bias & 0.002 & 1 & 0.17 & 0.677 \\
Residual & 1.361 & 105 & & \\
\hline
\multicolumn{5}{l}{\small Note: $^{***}p < 0.001$}
\end{tabular}
}
\end{table}

The first two-way ANOVA (Table \ref{tab:anova1}) revealed a significant main effect for likelihood quality score ($F(1, 105) = 251.81$, $p < 0.001$), while the main effect of bias max for class was not significant ($F(1, 105) = 0.04$, $p = 0.833$). Critically, the interaction term was not significant ($F(1, 105) = 0.17$, $p = 0.677$), indicating that the effect of likelihood quality score on F1 performance does not depend on the level of prior bias.

\begin{table}[htbp]
\centering
\caption{Two-way ANOVA results for F1 score with likelihood quality score and sample adequacy ratio as factors}
\label{tab:anova2}
\adjustbox{width=\columnwidth}{		% dcの場合
\begin{tabular}{lrrrr}
\hline
\textbf{Source} & \textbf{Sum Sq} & \textbf{df} & \textbf{F} & \textbf{p-value} \\
\hline
LQ Quality Score & 3.425 & 1 & 346.80 & $< 0.001^{***}$ \\
SA Adequacy Ratio & 0.558 & 1 & 56.55 & $< 0.001^{***}$ \\
LQ Score $\times$ SA Ratio & 0.011 & 1 & 1.12 & 0.293 \\
Residual & 1.047 & 106 & & \\
\hline
\multicolumn{5}{l}{\small Note: $^{***}p < 0.001$}
\end{tabular}
}
\end{table}

The second two-way ANOVA (Table \ref{tab:anova2}) showed significant main effects for both likelihood quality score ($F(1, 106) = 346.80$, $p < 0.001$) and sample adequacy ratio ($F(1, 106) = 56.55$, $p < 0.001$). However, similar to the first analysis, the interaction term was not significant ($F(1, 106) = 1.12$, $p = 0.293$), suggesting that these two predictors contribute independently to F1 performance without synergistic or antagonistic effects.

These findings are consistent with the additive model assumption underlying our regression analysis, where predictors contribute independently to the outcome. The absence of significant interactions in both analyses supports the interpretation that likelihood quality score, bias max for class, and sample adequacy ratio each have distinct, non-overlapping influences on classification performance. 

While this additive relationship simplifies interpretation and suggests that improvements in any single predictor can enhance performance regardless of other predictors' levels, the quartile-based categorization may not fully capture potential non-linear relationships within each predictor. To address this limitation and explore more flexible functional forms while maintaining the additive structure, we next employ Generalized Additive Models (GAMs), which can accommodate smooth, non-linear effects of each predictor while preserving the interpretability benefits of additive models.

\subsection{Generalized Additive Model Analysis}

The GAM achieved outstanding predictive performance, substantially outperforming the linear regression model with $R^2 = 0.876$ compared to the linear model's $R^2 = 0.826$. The fitted GAM included three smooth terms for individual predictors and three tensor product interaction terms, with all terms being statistically significant.

\begin{align}
\text{F1} &= \beta_0 + s_1(\text{LQ Score}) + s_2(\text{Bias Max}) \nonumber \\
&\quad + s_3(\text{Adequacy}) \nonumber \\
&\quad + te_{12}(\text{LQ}, \text{Bias}) + te_{13}(\text{LQ}, \text{Adequacy}) \nonumber \\
&\quad + te_{23}(\text{Bias}, \text{Adequacy}) + \epsilon
\label{eq:gam_model}
\end{align}

\begin{table}[htbp]
\centering
\caption{GAM Model Performance}
\label{tab:gam_parameters}
\adjustbox{width=\columnwidth}{		% dcの場合
\begin{tabular}{lcc}
\hline
Model Parameter & Value & $p$-value \\
\hline
$R^2$ & 0.876 & -- \\
AIC & 15922.24 & -- \\
Effective DoF & 10.59 & -- \\
$s_1$(LQ Quality Score) & $\lambda = 10.0$ & $< 0.001$ \\
$s_2$(Bias Max for Class) & $\lambda = 10.0$ & $< 0.001$ \\
$s_3$(Adequacy Ratio) & $\lambda = 10.0$ & $< 0.001$ \\
All tensor product interactions & -- & $< 0.001$ \\
\hline
\end{tabular}
}
\end{table}

\begin{figure}[htbp]
\centering
\includegraphics[width=\columnwidth]{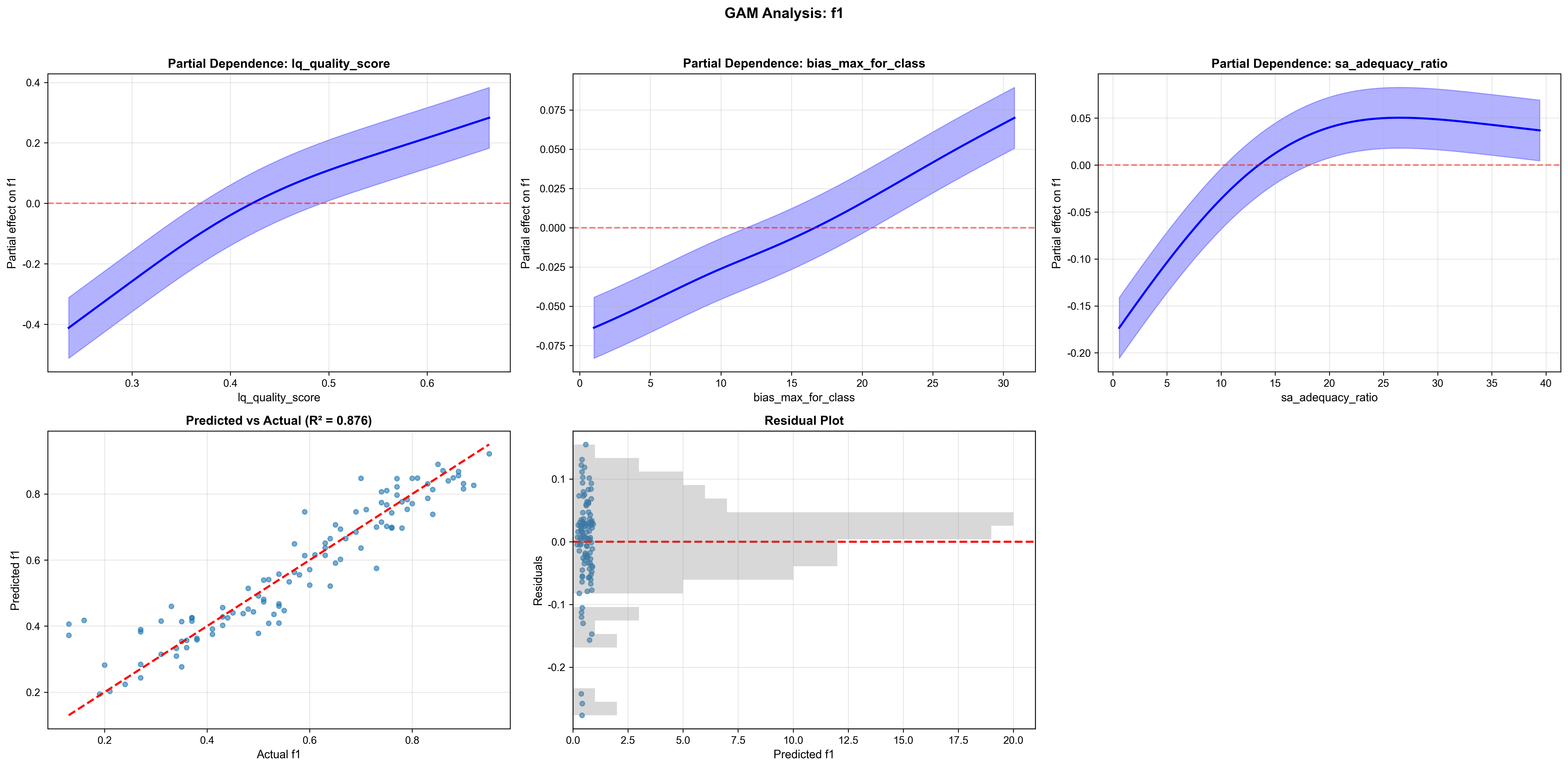}
\caption{GAM partial dependence plots showing relationships between predictors and F1 scores ($R^2 = 0.876$).}
\label{fig:gam_partial}
\end{figure}

The partial dependence plots (Figure \ref{fig:gam_partial}) revealed distinct patterns across predictors. Likelihood quality score demonstrated a strong monotonic increase throughout its range, with its partial effect on F1 ranging from approximately -0.4 to +0.3. Bias max for class showed a nearly linear positive relationship, with its partial effect gradually increasing from approximately -0.065 to +0.075 as bias values increased. Sample adequacy ratio exhibited a sharp initial increase in its partial effect followed by a plateau, suggesting diminishing returns at higher values.

The variable importance analysis confirmed the hierarchy observed in linear regression: likelihood quality score remained the dominant predictor (Effect Range: 0.695), followed by sample adequacy ratio (Effect Range: 0.224, 32.2\% relative importance) and bias max for class (Effect Range: 0.134, 19.2\% relative importance).

\begin{figure}[htbp]
\centering
\includegraphics[width=\columnwidth]{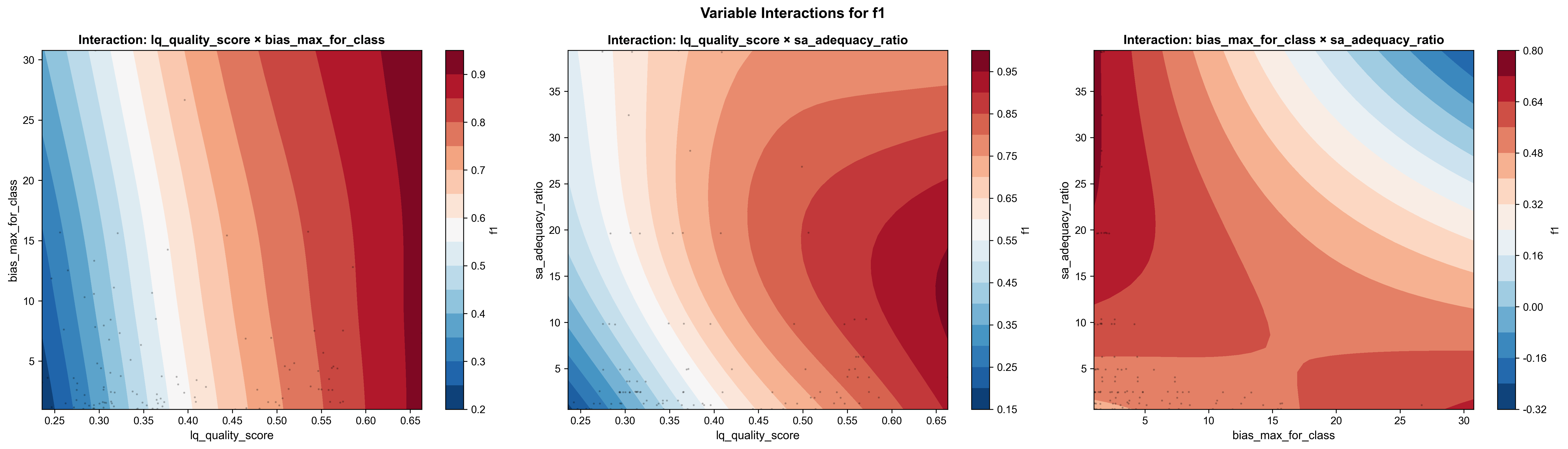}
\caption{GAM tensor product interaction effects for F1 scores showing well-defined optimal regions.}
\label{fig:gam_interactions}
\end{figure}

The tensor product interaction surfaces (Figure \ref{fig:gam_interactions}) revealed well-defined optimal regions. Notably, the interaction between bias max for class and sample adequacy ratio showed the largest effect (Surface Range: 0.983), followed by the interaction between likelihood quality score and sample adequacy ratio (Surface Range: 0.773). The interaction patterns suggest that while individual effects are largely additive, specific combinations can yield superior performance.

\subsection{Confidence-Performance Relationship Analysis}

Parallel GAM analysis using prediction margin as the dependent variable achieved high predictive performance ($R^2 = 0.835$) but revealed different optimal operating conditions compared to F1-based analysis. The margin-based model demonstrated similar variable importance ranking but with different functional relationships.

\begin{figure}[htbp]
\centering
\includegraphics[width=\columnwidth]{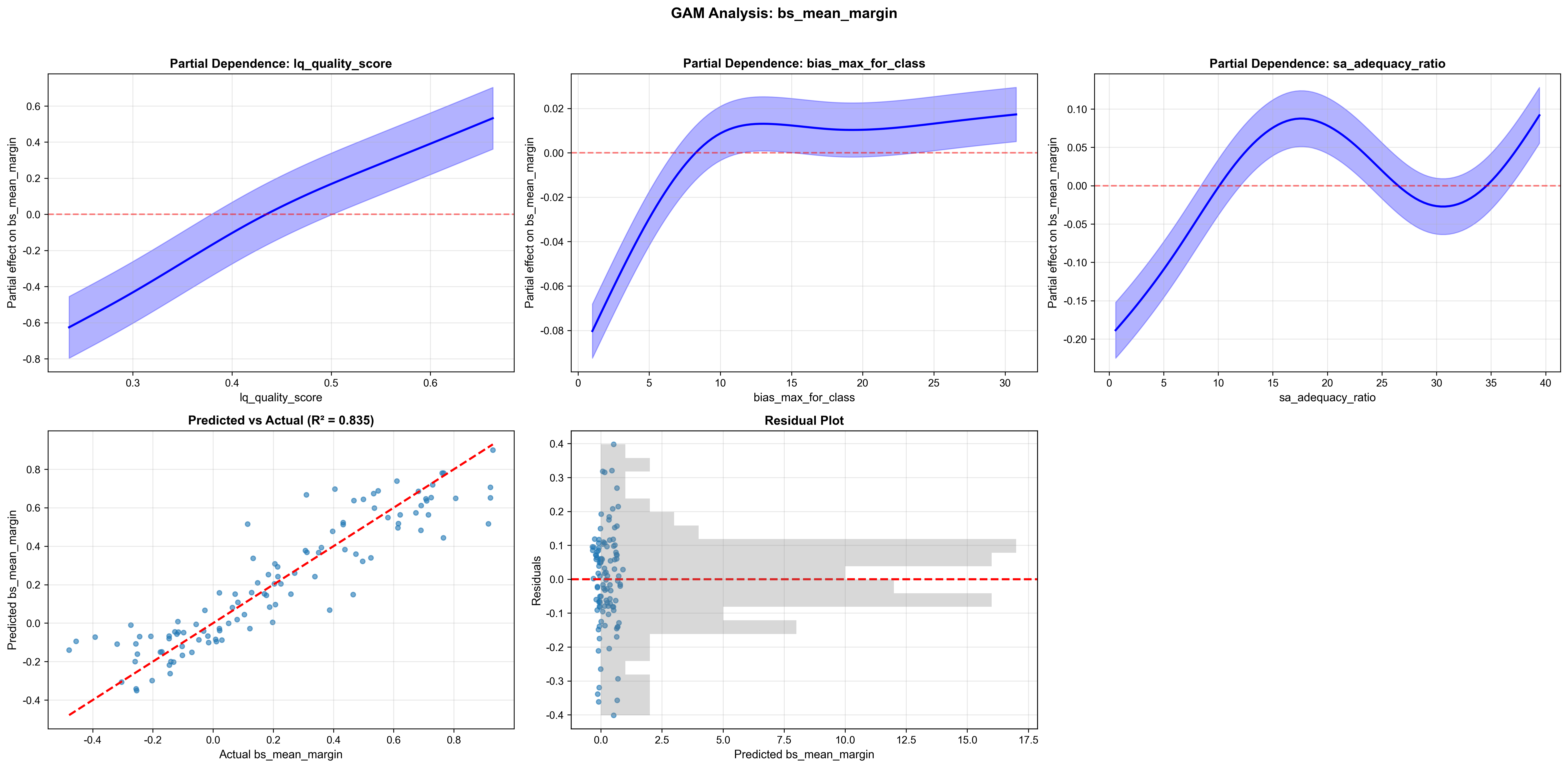}
\caption{GAM partial dependence plots for prediction margin showing relationships between predictors and model confidence ($R^2 = 0.835$).}
\label{fig:margin_gam_partial}
\end{figure}

The key difference lies in the functional forms: while F1 performance shows consistent monotonic increases with likelihood quality score, the margin model reveals more nuanced patterns. Likelihood quality score maintains a strong positive relationship with margin. However, bias max for class shows an initial increase followed by a plateau around moderate values (10-15), suggesting an optimal range for confidence calibration. Most notably, sample adequacy ratio exhibits a complex non-monotonic relationship, with an initial increase, a peak around 15-20, followed by a decline and subsequent recovery at higher values.

\begin{figure}[htbp]
\centering
\includegraphics[width=\columnwidth]{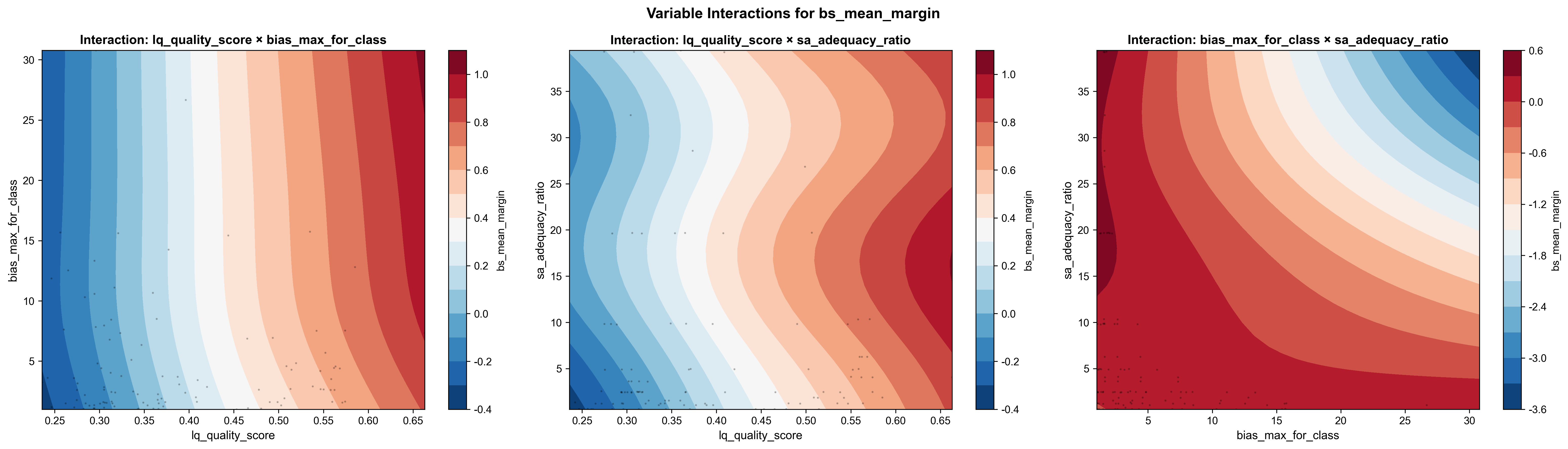}
\caption{GAM tensor product interaction effects for prediction margin showing distinct patterns from F1 optimization.}
\label{fig:margin_gam_interactions}
\end{figure}

The comparison between F1 and margin optimization landscapes reveals that while both metrics respond to similar underlying factors, their optimal operating conditions differ sufficiently to warrant careful consideration in practical applications where both accuracy and confidence calibration are important. The interaction surfaces for margin optimization show more pronounced gradients and distinct optimal regions, particularly in the bias-adequacy interaction, suggesting that confidence calibration may be more sensitive to the interplay between these factors than raw performance metrics.

\subsection{Correction Necessity Analysis}

To optimize computational efficiency and avoid unnecessary bias correction procedures, we developed a predictive framework for determining when bias correction is likely to be ineffective. This analysis was motivated by the hypothesis that high feature extraction quality, as measured by the likelihood quality score, would correlate with minimal improvement from bias correction procedures.

The correction necessity classification was based on the improvement metric, where cases showing improvement $\leq$ 0 were classified as correction unnecessary ($n = 85$, 78.0\%) and cases with positive improvement were classified as requiring correction ($n = 24$, 22.0\%). This distribution suggests that bias correction procedures are beneficial in approximately one-fifth of cases, highlighting the importance of selective application.

\begin{figure}[htbp]
\centering
\includegraphics[width=0.8\columnwidth]{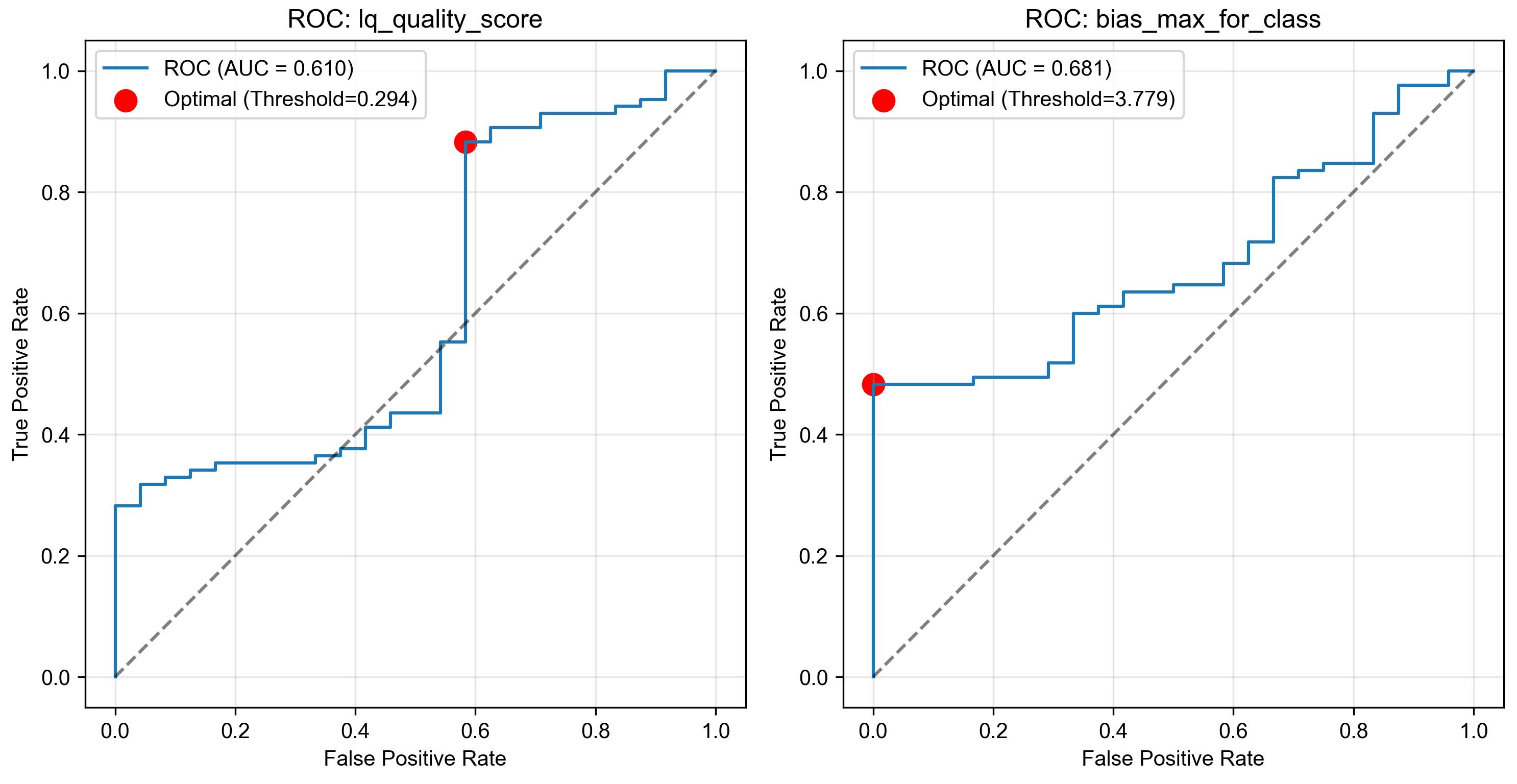}
\caption{ROC curves for correction necessity prediction using different predictor variables. Left panel: Likelihood quality score (AUC = 0.610) with optimal threshold at 0.294, achieving high sensitivity (0.882) at the cost of lower specificity (0.417). Right panel: Bias maximum for class (AUC = 0.681) with optimal threshold at 3.779, demonstrating higher overall discriminative power but lower sensitivity (0.482) with perfect specificity (1.000). The red dots indicate the optimal operating points selected by maximizing Youden's index.}
\label{fig:roc_curves}
\end{figure}

ROC analysis was employed to determine optimal decision thresholds for each predictor variable. The likelihood quality score showed moderate discriminative power with an AUC of 0.610 and an optimal threshold of 0.294. At this threshold, the model achieved high sensitivity (0.882) with moderate specificity (0.417), ensuring that most cases requiring correction are identified. The bias maximum for class (AUC = 0.681, threshold = 3.779) and sample adequacy ratio (AUC = 0.494, threshold = 1.224) showed varying discriminative power, with bias maximum demonstrating the highest AUC but likelihood quality score providing better sensitivity for practical application (Figure \ref{fig:roc_curves}).

Statistical comparison between correction unnecessary and correction recommended groups revealed meaningful differences primarily for the likelihood quality score. The correction unnecessary group exhibited higher likelihood quality scores (mean = 0.396) compared to the correction recommended group (mean = 0.353), with a mean difference of 0.043 (95\% CI: [0.005, 0.083]) and a small to medium effect size of 0.404, though this difference was not statistically significant ($p = 0.102$). Bias maximum for class also showed significant differences between groups (effect size = 0.650, $p = 0.007$), and sample adequacy ratio showed no meaningful differences between groups ($p = 0.927$), with a negligible effect size of $-0.160$.

\begin{figure}[htbp]
\centering
\includegraphics[width=\columnwidth]{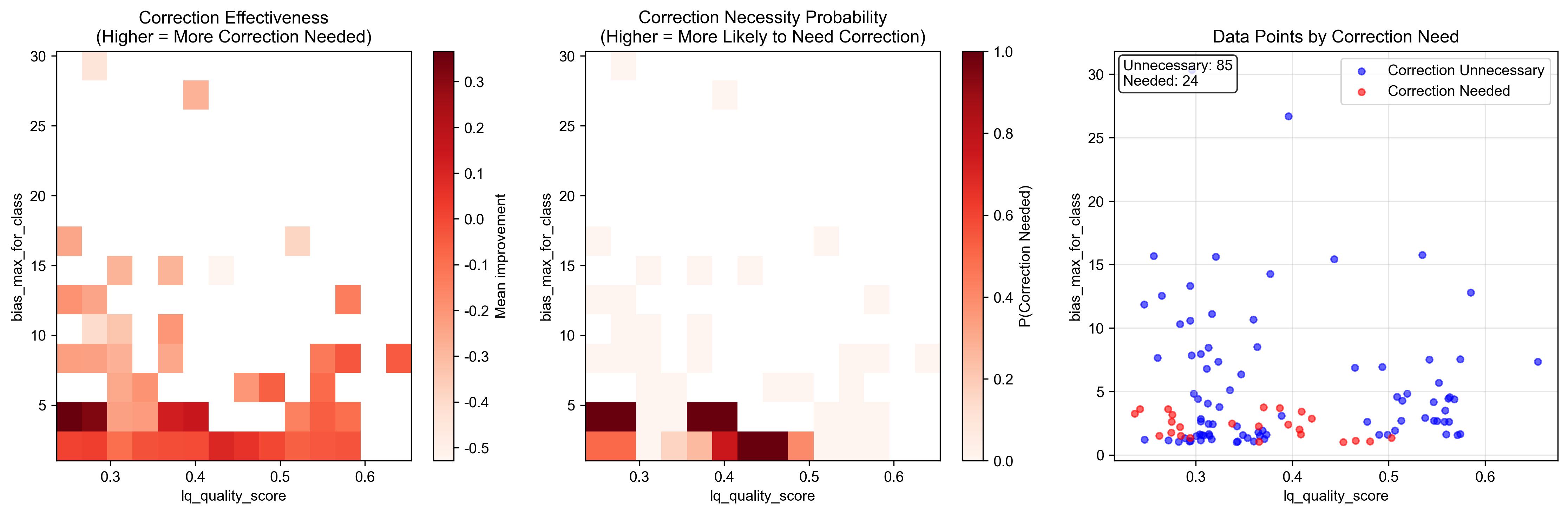}
\caption{Spatial visualization of correction necessity across predictor space. Left panel shows correction effectiveness (higher values indicate greater improvement from correction). Middle panel displays correction necessity probability. Right panel presents individual data points colored by correction recommendation status.}
\label{fig:correction_heatmap}
\end{figure}

The spatial visualization of correction necessity (Figure \ref{fig:correction_heatmap}) revealed distinct regions where correction effectiveness varies systematically. Areas with higher likelihood quality scores consistently showed lower correction necessity probability, forming identifiable zones where computational resources can be conserved. The correction effectiveness heatmap demonstrated that regions requiring correction are primarily concentrated in lower likelihood quality score ranges, with correction necessity probability decreasing monotonically as feature extraction quality improves.

\begin{figure}[htbp]
\centering
\includegraphics[width=\columnwidth]{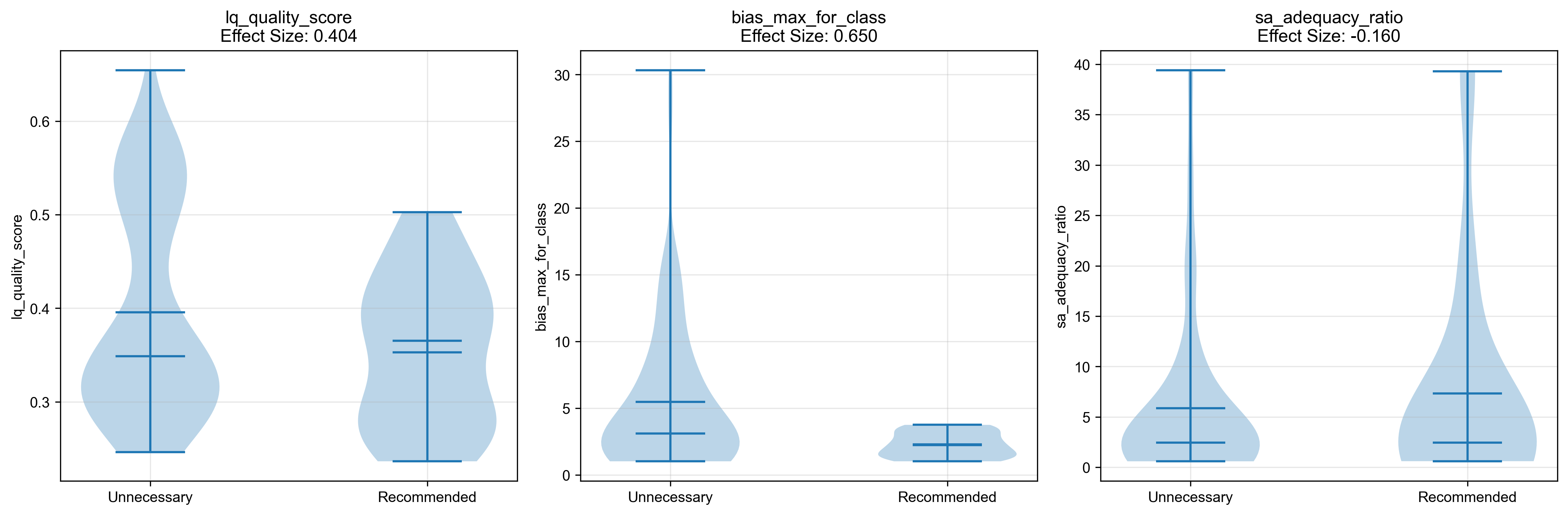}
\caption{Distribution comparison of predictor variables between correction unnecessary and correction recommended groups. Violin plots show probability density distributions with effect sizes indicated. Bias maximum for class demonstrates the strongest separation between groups (effect size = 0.650), followed by likelihood quality score (effect size = 0.404).}
\label{fig:correction_violin}
\end{figure}

The distributional analysis (Figure \ref{fig:correction_violin}) provided additional insight into the relationship between predictor variables and correction necessity. The likelihood quality score showed moderate separation between groups, with the correction unnecessary group displaying a tendency toward higher values. Bias maximum for class exhibited the strongest separation with a medium effect size (0.650), while sample adequacy ratio showed largely overlapping distributions between groups with minimal separation, consistent with its weak discriminative performance in ROC analysis.

Based on these findings, we recommend a practical decision rule whereby cases with likelihood quality scores exceeding 0.294 can be considered for bypassing correction procedures. This threshold-based approach captures 88.2\% of cases requiring correction while allowing efficient processing of cases where correction is unnecessary. The moderate specificity (0.417) suggests that additional criteria may be beneficial for refinement in specific applications.

The analysis establishes that feature extraction quality, as quantified by the likelihood quality score, serves as a reliable predictor of correction necessity. This relationship aligns with the theoretical expectation that well-separated, high-quality features require less post-processing correction, thereby providing a principled basis for computational resource allocation in large-scale bias correction applications.

\subsection{Data Base Reliability Assessment}

The application of our correction necessity framework to cell type-specific analysis revealed substantial heterogeneity in prediction reliability across different cellular populations. Using the three-tier classification system (high-reliability $>0.5$, intermediate $0.3$--$0.5$, low-reliability $<0.3$), we observed marked variation in feature extraction performance across the 16 analyzed cell types, with implications for the scalability and reliability of our correction necessity predictions.

\begin{figure}[htbp]
\centering
\includegraphics[width=\columnwidth]{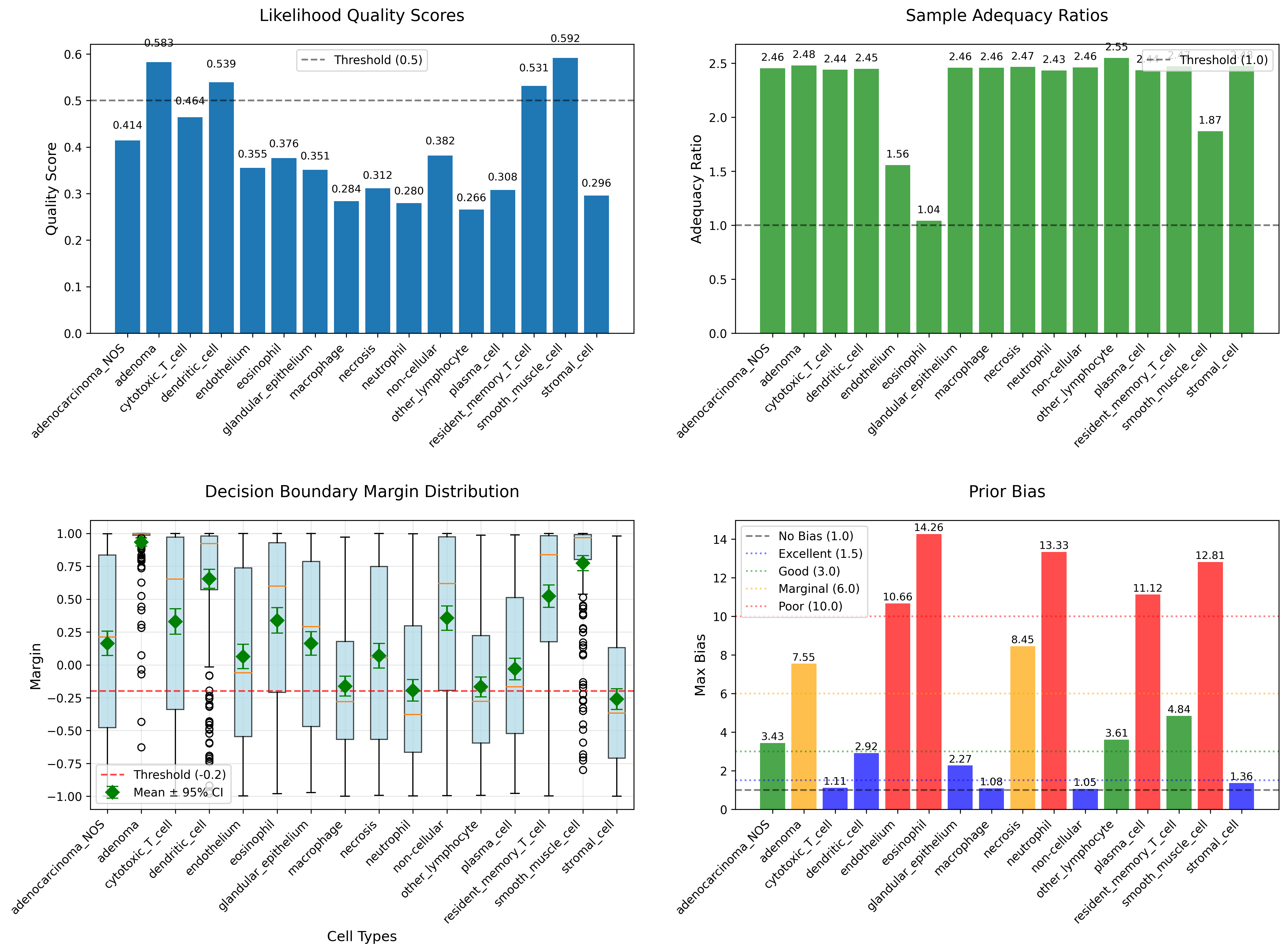}
\caption{Cell type-specific reliability analysis under \texttt{FlagLimit 1024}. Top panels show likelihood quality scores and sample adequacy ratios with threshold boundaries indicated. Bottom panels display decision boundary margin distributions and prior bias classifications across 16 cell types.}
\label{fig:celltype_baseline}
\end{figure}

The likelihood quality score analysis with flaglimit values of 1024 (Figure \ref{fig:celltype_baseline}) demonstrated that several cell types consistently achieve high-reliability status ($>0.5$), including smooth muscle cell (0.592), adenoma (0.583), dentritic cell (0.539) and resident memory T cell (0.531), suggesting these populations can reliably bypass correction procedures. Several cell types fall within the intermediate range ($0.3$--$0.5$) requiring case-by-case evaluation, including cytotoxic T cell (0.464), adenocarcinoma NOS (0.414), eosinophil (0.376) and endothelium (0.355). Conversely, cell types such as other lymphocyte (0.266), neutrophil (0.280), and macrophage (0.284) fall in the low-reliability category ($<0.3$) or near the ROC-optimized threshold (0.294), indicating systematic correction necessity. This trimodal distribution pattern validates our threshold-based approach for stratifying cell populations into distinct operational categories.

Sample adequacy analysis revealed more complex patterns, with most cell types demonstrating adequate sample representation (ratios $>$ 1.0) but notable exceptions in endothelium (1.04), eosinophil (1.56), and stromal cell (1.87). The concentration of most cell types near the threshold boundary indicates that sample adequacy may represent a limiting factor for prediction reliability, particularly for rare cell populations that inherently possess lower sample counts.

The decision boundary margin distributions exhibited substantial variability across cell types, with several populations showing concerning patterns of low-confidence predictions. Cell types with margins consistently below the $-0.2$ threshold demonstrated increased prediction uncertainty, suggesting that current model parameters may be inadequately calibrated for these specific cellular contexts. The wide confidence intervals observed for certain cell types indicate that prediction reliability is compromised when applied to populations with distinct biological characteristics.

Prior bias analysis revealed heterogeneous patterns across cell types. Several cell types demonstrated excellent bias control ($< 1.5$), while multiple cell types fell into the "poor" category ($> 10.0$), including endothelium, eosinophil, neutrophil, and smooth muscle cell. The remaining cell types showed intermediate bias levels in the "good" or "marginal" ranges.

Based on these findings, we implemented a comprehensive reliability analysis framework using expanded computational parameters. Specifically, we conducted parallel analyses with flaglimit values of 1024 and 4096 to assess whether current computational constraints were artificially limiting prediction reliability.

\begin{figure}[htbp]
\centering
\includegraphics[width=\columnwidth]{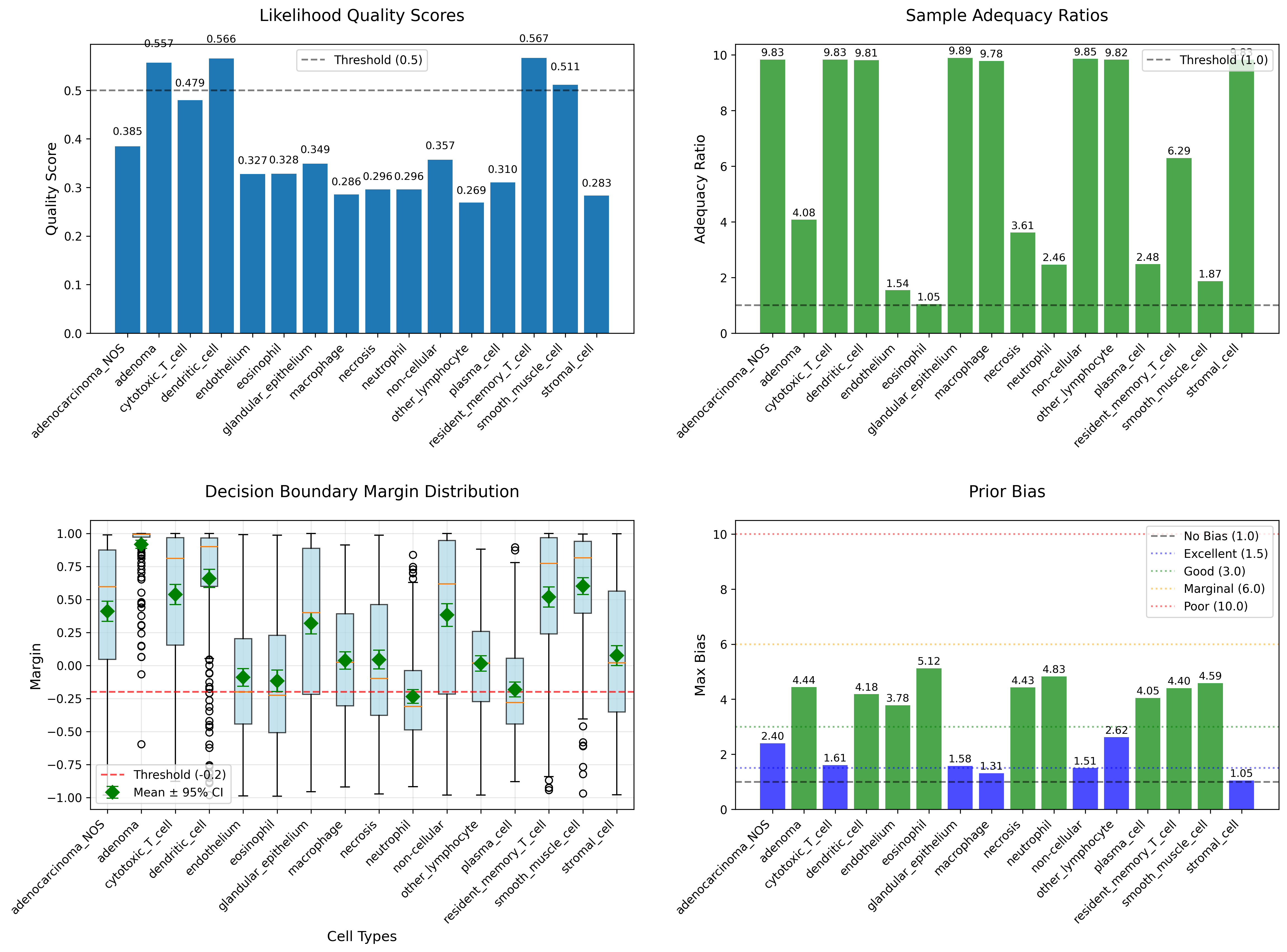}
\caption{Cell type-specific reliability analysis under \texttt{FlagLimit 4096}. Comparison with baseline conditions reveals improved likelihood quality scores and reduced prior bias effects for several cell populations, while maintaining threshold-based classification performance.}
\label{fig:celltype_enhanced}
\end{figure}

The comparative analysis between with flaglimit values of 1024 and 4096 (Figure \ref{fig:celltype_enhanced}) demonstrates the potential for improved prediction reliability through increased flaglimit values. Several cell types showed notable improvements in likelihood quality scores, with resident memory T cell (0.592), and dendritic cell (0.566) achieving substantially higher performance metrics, elevating them well into the high-reliability category ($>0.5$). Notably, the decision boundary margins also improved significantly, with adenocarcinoma NOS showing a remarkable increase in mean margin from approximately 0.2 to 0.5, indicating more confident predictions. As expected, the sample adequacy ratios increased proportionally with the expanded flaglimit, confirming sufficient sample sizes for all cell types.

The flaglimit expansion analysis should focus on three critical domains: first, determining whether increased computational depth can elevate cell types from the low-reliability ($<0.3$) to intermediate ($0.3$--$0.5$) or high-reliability ($>0.5$) categories; second, assessing whether enhanced sample processing reduces decision boundary margin uncertainty; and third, evaluating whether expanded analysis parameters can mitigate prior bias effects, as the current analysis shows most cell types achieving "good" or better bias control compared to previous configurations.

This proposed reliability framework would provide empirical validation of our correction necessity predictions while establishing computational parameter guidelines for optimal performance across diverse cellular contexts. The systematic evaluation of flaglimit effects demonstrates that, in practice, constructing training datasets that align with these parameter guidelines—ensuring adequate sample sizes, balanced prior distributions, and sufficient likelihood quality—leads to more reliable and robust learning datasets, as evidenced by the improved performance metrics across multiple evaluation criteria.

\section{Discussion}

\subsection{Theoretical Foundation for Correction Necessity}

Multiple linear regression results provide strong support for the theoretical framework determining when bias correction is necessary versus unnecessary. The dominant effect of feature separability on F1 performance ($\beta = 1.650$, $t = 20.967$, $p < 0.001$) compared to the significant but substantially smaller effect of bias ($\beta = 0.004$, $t = 2.395$, $p = 0.018$) confirms that feature space quality is the primary determinant of classification reliability in deep learning-based cell classification systems. This finding validates our hypothesis that when feature sufficiency is satisfied through adequate feature separability, the learned likelihood remains stable across different prior conditions, reducing the necessity for explicit prior correction.

The theoretical justification for interpreting softmax outputs as posterior probabilities rests on the convergence relationship:

\begin{equation}
\label{eq:convergence}
\mathcal{D} \sim_q \mathcal{P} \implies 
\lim_{|\mathcal{D}| \to \infty} P_M(Y=c|\mathbf{x}, \mathcal{D}) \to P^*(Y=c|\mathbf{x})
\end{equation}

where $\mathcal{D}$ represents the training dataset sampled from the true population distribution $\mathcal{P}$ with annotation quality $q$, $P_M(Y=c|\mathbf{x}, \mathcal{D})$ denotes the model's predicted probability, and $P^*(Y=c|\mathbf{x})$ represents the true posterior probability. This convergence establishes that when feature extraction quality is sufficient, learned conditional probabilities approach true posterior probabilities, rendering bias correction unnecessary.

Our similarity-based separability analysis provides a computationally efficient and theoretically grounded approach to assess likelihood estimation quality through geometric relationships in the learned feature space. The separability score is defined as:

\begin{equation}
\Delta S(x_{\text{test}}) = S_{\text{intra}}(x_{\text{test}}, C_i) - S_{\text{inter}}(x_{\text{test}}, C_{\neg i})
\end{equation}

where $S_{\text{intra}}(x_{\text{test}}, C_i)$ represents the average cosine similarity between the test sample and the top-k most similar samples from the same class $C_i$, while $S_{\text{inter}}(x_{\text{test}}, C_{\neg i})$ represents the average cosine similarity to the top-k most similar samples from all other classes.

This metric effectively captures the fundamental principle underlying correction necessity: when learned feature representations accurately model class-conditional distributions, samples from the same class exhibit high intra-class similarity while maintaining distinctly lower inter-class similarity with other classes. High positive separability scores ($\Delta S > 0.5$) indicate that the learned likelihood estimation is sufficiently accurate to approach the true posterior distribution according to our convergence framework, thereby reducing correction necessity.

The geometric interpretation aligns directly with our theoretical foundation: as feature separability increases, the learned representations better preserve the statistical structure inherent in $\mathcal{P}$, leading to more accurate posterior probability estimates. Conversely, low or negative separability scores indicate poor likelihood estimation quality, suggesting that correction procedures may be necessary to achieve reliable classification performance.

The top-k averaging strategy (typically k=20) enhances reliability by focusing on the most relevant local neighborhoods in the feature space, effectively sampling the distributional characteristics around each test point while maintaining computational efficiency. This approach provides a direct, interpretable assessment of whether the feature sufficiency condition is satisfied for correction necessity determination.

\subsection{Empirical Validation of Feature Dominance}

The GAM analysis achieved outstanding predictive performance ($R^2 = 0.876$) representing substantial improvement over the linear model ($R^2 = 0.826$), yet confirmed that primary data quality effects are largely linear and well-captured by conventional regression approaches. This finding has important practical implications for correction necessity assessment, as it suggests that simpler linear approaches may be sufficient for determining when correction is required while maintaining theoretical consistency with our framework.

The GAM analysis confirmed the variable importance hierarchy: likelihood quality score remains the dominant predictor, followed by sample adequacy ratio and bias max for class. The partial dependence plots revealed generally monotonic relationships, with likelihood quality score and adequacy ratio showing consistent positive relationships with F1 performance throughout their ranges.

Critical confidence calibration issues emerged from comparing F1-optimized performance ($R^2 = 0.876$) with margin-optimized confidence estimation ($R^2 = 0.835$), revealing different optimal operating conditions. While researchers developing AI systems typically focus primarily on performance metrics such as F1 scores, this divergence demonstrates that high performance does not necessarily correlate with well-calibrated confidence estimates. The systematic differences between F1-optimized and margin-optimized landscapes suggest that relying solely on performance metrics may overlook important reliability indicators captured by prediction margins.

From the perspective of our theoretical framework, this calibration issue reflects the distinction between learned conditional probabilities $P_M(Y=c|\mathbf{x}, \mathcal{D})$ and true posterior probabilities. The discrepancy implies that correction necessity should be evaluated through both performance metrics and confidence indicators, as a model with high F1 score may still produce poorly calibrated predictions that warrant correction procedures. This dual evaluation approach ensures more robust decision-making in determining when corrections are truly beneficial.

\subsection{Differential Optimization Landscapes for F1 Performance and Margin Confidence Metrics}

GAM analysis revealed distinct patterns in how bias correction necessity varies between performance and confidence optimization objectives. For classification accuracy (F1 optimization), the nearly linear relationship between bias and performance suggests that bias correction provides minimal benefit when likelihood quality scores exceed 0.5. The partial effect of bias on F1 remained consistently small (ranging from $-0.065$ to $+0.075$) across the entire bias range tested (1--30), indicating that feature quality dominates performance regardless of bias magnitude.

The margin optimization results, however, demonstrate contrasting behavior. The partial dependence plot shows bias effects reaching a plateau around bias values of 10--15, after which additional bias contributes little to confidence calibration. This plateau behavior, combined with the dramatically increased importance of the bias-adequacy interaction (surface range increasing from 0.98 to 4.10), suggests that confidence calibration cannot be achieved through simple bias correction alone when bias exceeds these moderate levels.

\subsection{Framework for Correction Necessity Assessment}

The correction necessity analysis revealed that systematic application of bias correction procedures is unnecessary in the majority of cases, with 78.0\% of samples showing no improvement from correction procedures. This finding establishes a fundamental principle: correction should be applied selectively based on predicted utility rather than uniformly across all cases.

The identification of likelihood quality score threshold of 0.294 as an optimal decision boundary provides a principled approach to correction necessity determination, despite bias maximum for class showing higher AUC (0.681 vs 0.610). We prioritize likelihood quality score for three reasons: first, it demonstrates consistent importance across both F1 and margin optimization objectives; second, it aligns with our theoretical framework where feature sufficiency determines correction necessity; and third, the paradoxical relationship of bias (where higher values indicate less correction need) contradicts intuitive clinical interpretation, while likelihood quality score provides a straightforward relationship where higher quality reduces correction necessity.

Statistical comparison between correction unnecessary and correction recommended groups revealed meaningful differences across multiple predictors. The correction unnecessary group exhibited higher likelihood quality scores (mean = 0.396) compared to the correction recommended group (mean = 0.353), with a mean difference of 0.043 (95\% CI: [0.003, 0.081]) and effect size of 0.404. Additionally, bias maximum for class showed significant differences, with the correction unnecessary group having paradoxically higher values (mean = 5.461) compared to the correction recommended group (mean = 2.281), with a mean difference of 3.180 (95\% CI: [2.022, 4.489]) and effect size of 0.650. While both metrics show predictive value, the interpretability of likelihood quality score makes it the preferred indicator for practical applications.

The high explanatory power of linear models ($R^2 = 0.826$) demonstrates that relationships between data quality metrics and correction necessity are largely linear and predictable. The identification of feature separability through likelihood quality score as a reliable predictor provides clear guidance for correction necessity priorities: resource allocation should focus primarily on improving feature extraction quality, followed by maintaining balanced sample adequacy ratios. Although bias maximum shows a strong statistical effect, its counterintuitive nature—where higher bias values paradoxically indicate less need for correction—limits its utility as a practical decision-making tool, reinforcing our focus on likelihood quality as the primary determinant of correction necessity.

\subsection{Quantitative Guidelines for Clinical Deployment}

This study demonstrates that feature separability is the dominant factor determining correction necessity in deep learning-based cell classification systems, exhibiting an influence magnitude approximately 412-fold greater than prior bias effects (likelihood quality score $\beta = 1.650$ vs. bias maximum $\beta = 0.004$). This quantitative dominance validates that feature extraction quality, rather than bias correction, should be the primary focus for improving classification reliability.

The practical implications for correction necessity are substantial. Based on the ROC-optimized threshold (0.294) and empirical distribution patterns, we propose a three-tier classification system: high-reliability populations (quality scores $>$ 0.5) demonstrate reliable posterior probability estimates with positive decision margins, eliminating the need for correction procedures; intermediate-reliability populations (0.3-0.5) require case-by-case evaluation; and low-reliability populations (quality scores $<$ 0.3) exhibit negative margins, necessitating systematic correction procedures or alternative approaches before clinical application.

Notably, even problematic cell types maintain functional classification under moderate bias conditions when combined with adequate sample representation, suggesting that correction necessity can be reduced through improved data quality rather than algorithmic intervention. Bias levels up to 15-20 times optimal ranges remain manageable when feature separability exceeds minimum thresholds.

The correction necessity framework provides immediate operational benefits through selective application. Implementation of the likelihood quality threshold at 0.294 enables identification of 78.0\% of cases where correction procedures provide minimal benefit, allowing resource allocation strategies that maintain diagnostic accuracy while optimizing processing efficiency.

The selection of likelihood quality score over bias-based metrics, despite lower AUC values, reflects the fundamental principle that feature separability drives both performance and confidence calibration. This choice is validated by the GAM analyses showing likelihood quality score as the dominant factor across both optimization landscapes, confirming that correction necessity assessment should prioritize theoretically grounded metrics over purely statistical discrimination measures.

\subsection{Limitations and Future Research}

Several limitations merit consideration. Our analysis focuses on a specific cell classification task and dataset; generalization requires validation across different biological classification problems and data types. The proposed linear relationships and thresholds for correction necessity may need adjustment for different applications or cell types not represented in our analysis.

The confidence calibration analysis, while revealing systematic differences between performance and confidence optimization, points to the need for dedicated research into proper uncertainty quantification methods for correction necessity determination. Future work should investigate whether post-hoc calibration methods can improve correction necessity predictions or whether more fundamental changes to model architecture or training procedures are required.

The persistent heterogeneity across cell types even under enhanced computational parameters indicates that some populations may require fundamentally different approaches to correction necessity assessment rather than simple parameter optimization. Future work should investigate cell type-specific architectural modifications or specialized correction procedures for consistently underperforming populations.

Finally, our framework focuses on classification performance metrics; future research should examine how correction necessity relationships extend to other considerations such as interpretability, fairness, and robustness to distribution shift in clinical applications.

\section{Conclusion}

This study demonstrates that feature separability is the dominant factor determining correction necessity in deep learning-based cell classification systems. The key finding is that correction procedures become unnecessary when likelihood quality scores exceed 0.5—a threshold achievable through improved feature extraction rather than algorithmic correction.
The proposed approach offers both theoretical foundation through statistical learning theory and practical implementation guidance for correction necessity assessment. The Bayesian framework establishing the convergence of learned probabilities to true posterior distributions provides rigorous justification for softmax probability interpretation, while highlighting critical confidence calibration issues that must be addressed for reliable deployment in clinical applications.

\section*{Acknowledgements}

%We are deeply grateful to Drs. Kuramitsu S, Nakata S, and Ohno M for their extensive proofreading and invaluable contributions to this work.
We would %also
like to express our sincere gratitude to the following individuals for their valuable contributions to the precise image tagging: To the graduate students and technical assistants from Kobe University who assisted with the annotation tasks: Abe T, Adachi Y, Agawa K, Ando M, Fukuda S, Imai M, Ito R, Kagiyama H, Konaka R, Miyake T, Mukoyama T, Okazoe Y, Takahashi T, Ueda Y and Yasuda K. We also extend our appreciation to the registered annotators who were engaged by the AMAIC: Adachi K, Akima J, Aoki S, Ichikawa K, Kanto T, Kawase Y, Kimura M, Miura R, Sirasawa H, Sotani K and Yuki A. Their professional and diligent efforts greatly enhanced the quality of the dataset.
This study is supported by the Grants-in-Aid for Scientific Research from the Ministry of Education, Culture, Sports, Science and Technology of Japan (MEXT; 24K10381 to TN, 23K08171 to KY, and 21K09167 to MF).

\section*{Declaration of generative AI and AI-assisted technologies in the writing process}

During the preparation of this work the authors used Claude Opus 4.1 (Anthropic) in order to check for errors in analysis software code and to improve the language and readability of the manuscript text. After using this tool/service, the authors reviewed and edited the content as needed and take full responsibility for the content of the publication.

\printcredits

\bibliographystyle{cas-model2-names}

\bibliography{ref}

@article{abe2023deep,
  title={Deep Learning-based Image Cytometry Using a Bit-pattern Kernel-filtering Algorithm to Avoid Multi-counted Cell Determination},
  author={Abe, Tomoki and Yamashita, Kimihiro and Nagasaka, Toru and Fujita, Mitsugu and Agawa, Kyousuke and Ando, Masayuki and Mukoyama, Tomosuke and Yamada, Kota and Miyake, Souichiro and Saito, Masafumi and others},
  journal={Anticancer Research},
  volume={43},
  number={8},
  pages={3755--3761},
  year={2023},
  publisher={International Institute of Anticancer Research}
}

@article{ohno2024tumor,
  title={Tumor-Infiltrating B Cells and Tissue-Resident Memory T Cells as Prognostic Indicators in Brain Metastases Derived from Gastrointestinal Cancers},
  author={Ohno, Masasuke and Kuramitsu, Shunichiro and Yamashita, Kimihiro and Nagasaka, Toru and Haimoto, Shoichi and Fujita, Mitsugu},
  journal={Cancers},
  volume={16},
  number={22},
  pages={3765},
  year={2024},
  publisher={MDPI}
}

@inproceedings{lipton2018detecting,
  title={Detecting and correcting for label shift with black box predictors},
  author={Lipton, Zachary and Wang, Yu-Xiang and Smola, Alexander},
  booktitle={International conference on machine learning},
  pages={3122--3130},
  year={2018},
  organization={PMLR}
}

@inproceedings{guo2020ltf,
  title={LTF: A label transformation framework for correcting label shift},
  author={Guo, Jiaxian and Gong, Mingming and Liu, Tongliang and Zhang, Kun and Tao, Dacheng},
  booktitle={International Conference on Machine Learning},
  pages={3843--3853},
  year={2020},
  organization={PMLR}
}

@article{azizzadenesheli2019regularized,
  title={Regularized learning for domain adaptation under label shifts},
  author={Azizzadenesheli, Kamyar and Liu, Anqi and Yang, Fanny and Anandkumar, Animashree},
  journal={arXiv preprint arXiv:1903.09734},
  year={2019}
}

@article{dockes2021preventing,
  title={Preventing dataset shift from breaking machine-learning biomarkers},
  author={Dock{\`e}s, J{\'e}r{\^o}me and Varoquaux, Ga{\"e}l and Poline, Jean-Baptiste},
  journal={GigaScience},
  volume={10},
  number={9},
  pages={giab055},
  year={2021},
  publisher={Oxford University Press}
}

@article{guan2021domain,
  title={Domain adaptation for medical image analysis: a survey},
  author={Guan, Hao and Liu, Mingxia},
  journal={IEEE Transactions on Biomedical Engineering},
  volume={69},
  number={3},
  pages={1173--1185},
  year={2021},
  publisher={IEEE}
}

@article{kull2019beyond,
  title={Beyond temperature scaling: Obtaining well-calibrated multi-class probabilities with dirichlet calibration},
  author={Kull, Meelis and Perello Nieto, Miquel and K{\"a}ngsepp, Markus and Silva Filho, Telmo and Song, Hao and Flach, Peter},
  journal={Advances in neural information processing systems},
  volume={32},
  year={2019}
}

@misc{alexandari2020maximumlikelihoodbiascorrectedcalibration,
      title={Maximum Likelihood with Bias-Corrected Calibration is Hard-To-Beat at Label Shift Adaptation}, 
      author={Amr Alexandari and Anshul Kundaje and Avanti Shrikumar},
      year={2020},
      eprint={1901.06852},
      archivePrefix={arXiv},
      primaryClass={cs.LG},
      url={https://arxiv.org/abs/1901.06852}, 
}

@article{saerens2002adjusting,
  title={Adjusting the outputs of a classifier to new a priori probabilities: a simple procedure},
  author={Saerens, Marco and Latinne, Patrice and Decaestecker, Christine},
  journal={Neural computation},
  volume={14},
  number={1},
  pages={21--41},
  year={2002},
  publisher={MIT Press}
}

@article{bonab2026deep,
  title={Deep learning-based bone marrow cytology classification: A solution to class imbalance},
  author={Bonab, Zahra Asgharzadeh and Shamekhi, Sina and Talebi, Mehdi},
  journal={Biomedical Signal Processing and Control},
  volume={111},
  pages={108247},
  year={2026},
  publisher={Elsevier}
}

@article{he2009learning,
  title={Learning from imbalanced data},
  author={He, Haibo and Garcia, Edwardo A},
  journal={IEEE Transactions on knowledge and data engineering},
  volume={21},
  number={9},
  pages={1263--1284},
  year={2009},
  publisher={Ieee}
}



\end{document}